\documentclass[aps,pre,reprint,superscriptaddress, floatfix,longbibliography]{revtex4-1} %change to reprint to see two column format
\usepackage{graphicx}				% Use pdf, png, jpg, or eps§ with pdflatex; use eps in DVI mode
\usepackage{import}								% TeX will automatically convert eps --> pdf in pdflatex		
\usepackage{amssymb}
\usepackage{amsmath}
\usepackage{verbatim}
\usepackage{hyperref}
\DeclareMathOperator\erf{erf}
\usepackage{graphicx}
\graphicspath{ {images/} }
\usepackage{booktabs}
\usepackage{multirow}
\usepackage{array}

\begin{document}

%\title{Effect of dissociative search mechanism on successful search probabilities in DNA mismatch repair}
\title{Potential evolutionary advantage of a dissociative search mechanism in DNA mismatch repair}
\author{Kyle Crocker}
\affiliation{Department of Physics, The Ohio State University, Columbus, Ohio 43210, USA}  
\author{James London}
\affiliation{Department of Cancer Biology and Genetics, The Ohio State University, Columbus, Ohio 43210, USA}  
\author{Andr\'es Medina}
\affiliation{Department of Physics, The Ohio State University, Columbus, Ohio 43210, USA}  
\author{Richard Fishel}
\affiliation{Department of Cancer Biology and Genetics, The Ohio State University, Columbus, Ohio 43210, USA}
\author{Ralf Bundschuh}
\affiliation{Department of Physics, Department of Chemistry and Biochemistry, Division of Hematology, Department of Internal Medicine, The Ohio State University, Columbus, Ohio 43210, USA}
\date{\today}							% Activate to display a given date or no date

%%MORE AUTHORS?? who worked on the gillespie code previously?

\begin{abstract}

Protein complexes involved in DNA mismatch repair appear to diffuse along dsDNA in order to locate a hemimethylated incision site via a dissociative mechanism. Here, we study the probability that these complexes locate a given target site via a semi-analytic, Monte Carlo calculation that tracks the association and dissociation of the complexes. We compare such probabilities to those obtained using a non-dissociative diffusive scan, and determine that for experimentally observed diffusion constants, search distances, and search durations \textit{in vitro}, there is neither a significant advantage nor disadvantage associated with the dissociative mechanism in terms of probability of successful search, and that both search mechanisms are highly efficient for a majority of hemimethylated site distances. Furthermore, we examine the space of physically realistic diffusion constants, hemimethylated site distances, and association lifetimes and determine the regions in which dissociative searching is more or less efficient than non-dissociative searching. We conclude that the dissociative search mechanism is advantageous in the majority of the physically realistic parameter space.
 
\end{abstract}
 
 
 \maketitle
 
 \section{Introduction}
 
 DNA mismatch repair (MMR) is a molecular process by which errors in DNA sequence indicated by mismatched base pairs are corrected. Failure of this process is the cause of many cancers~\cite{martinlopez2013}, but a complete mechanistic description of the process does not yet exist~\cite{martinlopez2013, Reyes2015, Liu2016}. The MMR process is evolutionarily conserved from prokaryotes to eukaryotes~\cite{wang2006, iyer2006, Fishel2015}, so \textit{E. coli} MutS, MutL, and MutH proteins may be productively used to study MMR. In \textit{E. coli}, MMR consists of the following steps. First, MutS recognizes a mismatched site on a DNA strand and associates with the DNA. This MutS then binds MutL from solution, which in turn can bind MutH. MutH then nicks the newly synthesized, erroneous DNA strand. Excision, followed by polymerization and ligation, complete the repair process~\cite{iyer2006, Fishel2015}. 
 
 Here, we describe a quantitative model of the process by which the MMR proteins determine which strand is newly synthesized. Since \textit{E. coli} methylates its DNA strands whenever a GATC base sequence appears, a newly synthesized strand differs from existing strands in that it is not yet methylated. A MutL activated MutH, therefore, nicks the new strand at a hemimethylated site, and the strand containing the nick is excised. In order to create this nick, however, a hemimethylated site must first be recognized. The hemimethylated sites may be thousands of base pairs away from the mismatch (and therefore the place at which the MutS proteins bind to the DNA), so recognition of a hemimethylated site is not a trivial problem. Through single molecule probing of the MMR process \textit{in vitro}, Liu \textit{et al.} recently found extremely stable toroidal protein clamps diffusing along the DNA strand while transiently associating and dissociating from each other in order to reach and recognize a hemimethylated site~\cite{Liu2016}. It is this diffusion mechanism that is the subject of our quantitative model.
 
 Protein searches of DNA for specific sites are common, and searches involving non-toroidal proteins have been studied extensively: Berg \textit{et al.} derived a complete mathematical model of this search process in terms of association and dissociation rates, as well as geometrical considerations~\cite{berg1981}. Givaty \textit{et al.} developed a molecular simulation based on electrostatic forces of non-toroidal DNA binding proteins searching DNA and tracked non-toroidal protein motion. They found that the most efficient DNA searches consist of $\sim 20\%$ sliding and $\sim 80\%$ three dimensional diffusion~\cite{givaty2008}. The toroidal, ``sliding clamp" protein structure which we are interested in here was reported by O'Donnell \textit{et al.} in the context of an \textit{E. coli} polymerase, DNA polymerase III holoenzyme, which is stabilized on the DNA by the $\beta$-clamp that encircles the DNA~\cite{odonnell1992}. More recently, Daitchmen \textit{et al.} have used molecular dynamics simulations to study the diffusion of these protein clamps and report on the way in which the physical properties of the clamps affect the diffusion dynamics~\cite{daitchman2018}. However, all of the previous studies that we are aware of have focused on individual proteins rather than the search process as a whole.
 
 The focus of this paper is to quantitatively model the observed protein clamp association-dissociation mechanism present in MMR protein clamp diffusion. While this is similar to the non-toroidal search mechanism described by Berg in that it is characterized by a transition between a slow searching state and a fast non-searching state, the time distribution of the fast state is different in each case. In particular, the transition from the dissociated fast state into the slow searching state is governed by 3-D diffusion in the non-toroidal proteins discussed by Berg~\cite{berg1981}, whereas the toroidal structure of the proteins that we consider, while allowing dissociation of the proteins from each other, prevents release from the DNA and thus restricts their motion to a single dimension. This structure also prevents transfer between nearby DNA segments~\cite{odonnell1992, daitchman2018, Kong1992}. 
 
After construction of the quantitative model, we investigate if the association-dissociation mechanism serves to increase the efficiency with which a hemimethylated site is found, as compared to a more straightforward situation in which the proteins are unable to dissociate from one another (or, equivalently, there is only a single protein). If this were the case, it could provide an evolutionary pressure favoring the association-dissociation mechanism. We find that although the association-dissociation mechanism makes little difference at the observed \textit{E. coli} parameters, there is a much larger section of parameter space in which the association-dissociation mechanism is beneficial as opposed to detrimental when compared to a case in which the proteins do not dissociate.
 
This paper begins with a summary of the Liu \textit{et al.} experiments that led to our model, including a tabulation of experimental parameters relevant to the model in section~\ref{sec:experiment}. In section~\ref{sec:model}, the model itself is described both physically and mathematically. Section~\ref{sec:probability} presents our approach to calculating the probability of finding the hemimethylated site from the model. The main findings concerning the probability of finding the hemimethylated site are then presented in section~\ref{sec:results}, and finally the implications of those results are discussed in section~\ref{sec:conclusions}, along with potential future directions of research in this area. Several of the detailed derivations are relegated to various appendices.
 
\section{Experimental observation of dissociative search mechanism}\label{sec:experiment}
In this section, we briefly summarize the experimental observations by Liu \textit{et al.}~\cite{Liu2016} that underlie the model developed in this paper. Additionally, we compile in Table~\ref{table:expParams} experimentally determined quantities used to determine values of model parameters, since we refer to these quantities throughout the paper.

In the experiment by Liu \textit{et al.}, interactions of \textit{Escherichia coli} DNA mismatch repair proteins MutS, MutL, and MutH with dsDNA were imaged via TIRF microscopy. Of particular interest is what will be referred to as the dissociative search mechanism, so-called because of the many cycles in which MutS and MutL associate into a single complex and then dissociate into two separate complexes before re-forming a single complex as they diffuse along the DNA in order to locate the hemimethylated site. (We called this mechanism the ``association-dissociation mechanism" in the introduction for clarity, but for the remainder of the paper we switch to the less cumbersome ``dissociative mechanism.")

In particular, when MutS binds to a mismatch, it forms a stable clamp in the presence of ATP. It then diffuses along the DNA strand. MutL may then bind to MutS, forming a new clamp which together diffuses more slowly along the DNA. This slower diffusion implies frequent interaction with the DNA backbone, thus indicating that the MutS-MutL  clamp is capable of ``searching" the DNA for a hemimethlyated site~\cite{Liu2016}. Interestingly, MutL often dissociates from MutS, and the two proteins form two independent, stable, and freely diffusing clamps, each of which diffuses much more quickly than the MutS-MutL complex and is therefore not interacting with the DNA frequently enough to perform a search~\cite{Liu2016}. If the dissociated clamps diffuse back into a state in which they are adjacent along the DNA, they are able to reassociate and continue to search the DNA together.  Finally, MutH associates with MutL in order to cleave the newly synthesized DNA  strand at the hemimethylated site.  Measured association lifetimes and diffusion constants for the dissociative search are compiled in Table \ref{table:expParams}. Note that the diffusion of the MutS protein alone is $\sim 10$ times faster than the diffusion of the MutS-MutL complex, and that the diffusion of the MutL protein in the absence of MutH is a factor of $\sim 20$ faster than that of the MutS protein alone. In the presence of MutH, MutS and MutL diffuse at similar rates. Furthermore, the addition of MutH does not seem to have a significant effect on the MutS-MutL diffusion constant~\cite{Liu2016}.  

\begin{table}[ht]
\centering
\begin{tabular}{|>{\raggedright}p{.20\columnwidth}|l|p{.29\columnwidth}| >{\arraybackslash}p{.29\columnwidth}| }
%\begin{tabular}{ |l|l|l| }
\hline
Quantity & Symbol & SL Value & SLH Value  \\
\hline
Search complex diffusion constant & $D_{\mathrm{SL,M}}$  & $(6 \pm 3) \times 10^4$  bp$^2$/s & $(8 \pm 5)\times 10^4$  bp$^2$/s  \\
 \hline
 MutS diffusion constant & $D_{S}$ & $(7 \pm 2) \times 10^5$  bp$^2$/s & NA  \\
 \hline
 MutL diffusion constant & $D_{L}$ & $(1.4\pm0.6) \times 10^7$  bp$^2$/s & $(6 \pm 5) \times 10^5$ bp$^2$/s \\
 \hline
  MutS-MutL association lifetime & $\tau_{\mathrm{A,M}}$ & $30 \pm 3$ s & NA \\
 \hline
  MutS-DNA association lifetime & $\tau_{S}$ & $185 \pm 35$ s & NA \\
 \hline
 MutL-DNA association lifetime & $\tau_{L}$ & $850 \pm 150$ s & NA \\
 \hline
\end{tabular} \\
\caption{Summary of relevant quantities measured by Liu \textit{et al.}~\cite{Liu2016}. The column labelled ``SL Value'' gives the value of each quantity in the absence of MutH, whereas the column labelled ``SLH Value'' gives the value of each quantity in the presence of MutH. Note that some values were not measured separately in the presence of MutH (indicated by NA), so we will assume that MutH does not change these values. In Liu's experiment, $17.3$ kb of dsDNA was stretched over $4.4$ $\mu$m, so we use $4.4\textrm{ } \mu\mathrm{m} = 17.3\textrm{  kb}$ to convert from reported values in $\mu$m by Liu \textit{et al.} to units of bp, which is more convenient for our model.  }
\label{table:expParams}
\end{table}

The objective of this paper is to quantitatively study the effect of this dissociative mechanism on search efficiency. In particular, there are  two competing effects of dissociative diffusion on search efficiency that make its overall effect unclear. Since it makes the overall diffusion faster (compared to a system that always remains in the slow, searching state), it increases the region of the DNA that the protein clamps are able to visit. However, since proteins in the dissociated state are unable to actually search the DNA, the amount of DNA actually searched may decrease if the proteins do not reassociate often enough. 

\section{Model}\label{sec:model}
\begin{figure}
\centering
\includegraphics[width=0.6\columnwidth]{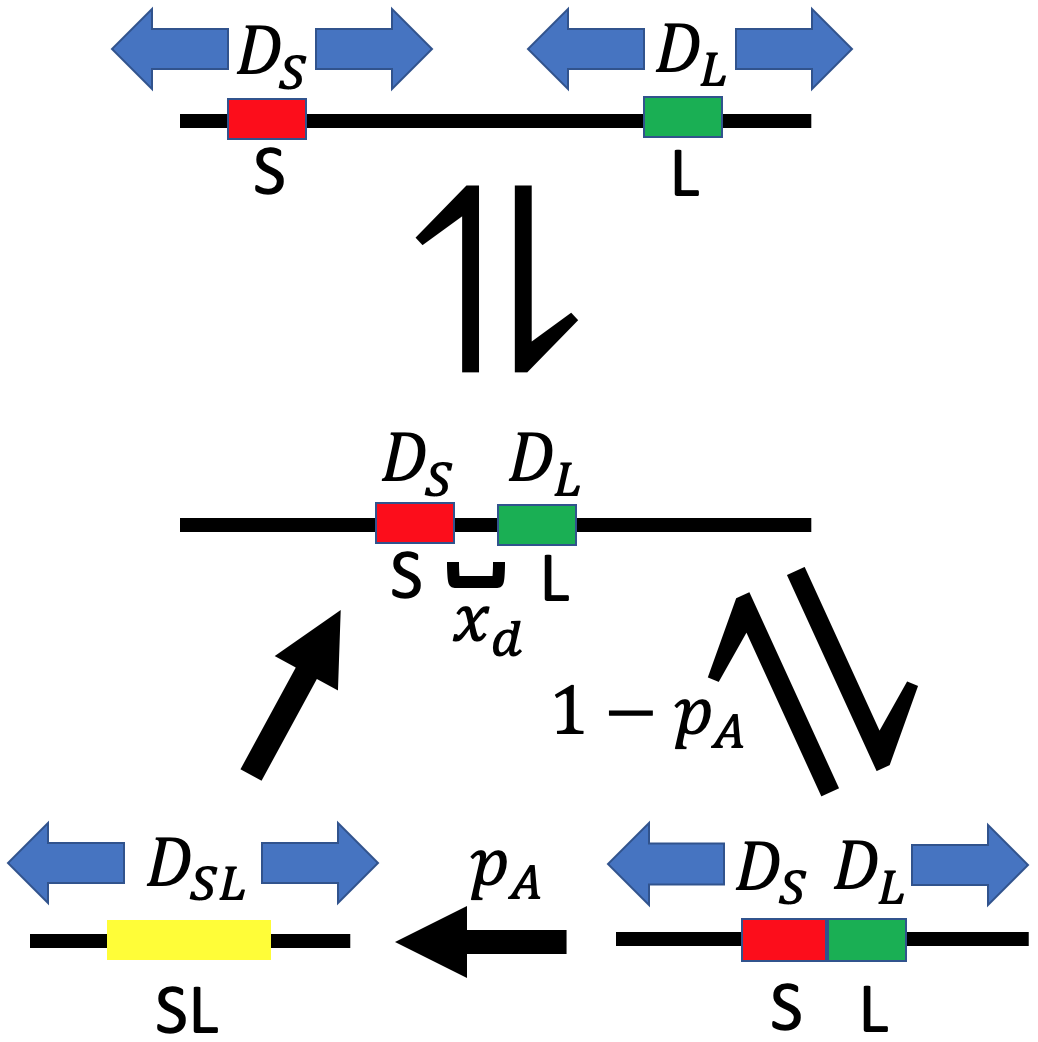}
\caption{(color online) Illustration of the model used to calculate successful hemimethylated site search probabilities in DNA mismatch repair. The DNA is modeled as a one dimensional track along which the proteins travel. MutS and MutL protein clamps diffuse along the DNA with rates specified by $D_S$ and $D_L$, respectively, until they are directly adjacent to each other. They may then either diffuse away from each other without associating or associate into a combined, searching MutS-MutL. These happen with probabilities $1 - p_A$ and $p_A$, respectively. The combined MutS-MutL can in turn diffuse along the DNA with a new rate specified by $D_\mathrm{SL}$, and is capable of searching the DNA over which it passes. After an average lifetime $\tau_{\mathrm{A},\mu}$, this complex dissociates into separate MutS and MutL clamps.}
\label{fig:modelSchematic}
\end{figure}

To determine the effect of the dissociative DNA mismatch repair search mechanism on search probabilities, we propose the microscopic model illustrated in Fig.\ref{fig:modelSchematic}. In the model, the search begins with an associated MutS-MutL protein complex. The complex then diffuses in one dimension along the DNA with diffusion constant $D_{\mathrm{SL},\mu}$. During this time, any portion of the DNA over which the complex passes is considered ``searched'' and the overall search is considered successful if a hemimethylated site is reached in this state.
The MutS-MutL complex dissociates with some average lifetime $\tau_{\mathrm{A},\mu}$ into independent MutS and MutL clamps initially separated by a distance $x_d$ with diffusion constants $D_S$ and $D_L$, respectively. The individual MutS and MutL clamps diffuse along the DNA until they come into contact again. Once in contact,  the proteins reassociate with each other with an association probability $p_A$ or continue independent diffusion starting from a distance $x_d$ with probability $1-p_A$. This dissociation-reassociation cycle continues until one or both of the proteins dissociate from the DNA. However, dissociation of the MutS and MutL clamps from the DNA is not modeled directly; instead, a ``cutoff time,'' based on the MutS association lifetime $\tau_S = 185$ s (which is shorter than the MutL lifetime and thus provides the more stringent cutoff), is used~\cite{Liu2016, Fishel2015} after which the search is declared unsuccessful. Since MutH binding to MutL changes the diffusion rates, we provide results that correspond to a search with non-MutH rates as well as results that correspond to a search with MutH rates. These two scenarios provide the two limiting cases, since it is not known if MutH tends to associate with MutL early or late in the search.

\section{Successful Hemimethylated Site Search Probability}\label{sec:probability}

Analysis of the dissociative search mechanism is performed in terms of successful hemimethylated site search probabilities. A successful search is defined as one in which the MutS-MutL complex visits the hemimethylated site before the cutoff time passes. These search probabilities are calculated for different hemimethylated site distances from the original MutS-MutL position. 

\subsection{Successful hemimethylated search probability of non-dissociative search}

The baseline to which we compare successful hemimethylated search probabilities using the dissociative search mechanism are the equivalent probabilities for searches in which the clamps remain associated with each other for the entire search (pure 1-dimensional diffusion), which will be referred to as ``non-dissociative" or ``purely diffusive" searches. The probability of a successful non-dissociative search can be derived analytically following Redner~\cite{redner2001}. We start with the diffusion equation in the case of a MutS-MutL clamp:
\begin{equation}
     \frac{\partial p \left (x, t | x_0 \right )}{\partial t} = D_{\textrm{SL}} \frac{\partial^2 p \left (x, t | x_0 \right )}{\partial x^2},
\end{equation}
where $D_{\textrm{SL}}$ is the diffusion constant associated with the MutS-MutL clamp and $x_0$ is the position along the DNA at which the non-dissociative search begins. $p(x, t) dx$ is the probability that the clamp will be searching position $x$ at time $t$. 

In order to consider the probability that the search reaches the hemimethylated site $x_{\textrm{meth}}$, we first solve this differential equation in the presence of an absorbing boundary condition at $x_{\textrm{meth}}$. Mathematically, this condition is expressed as $p(x_{\textrm{meth}}, t) = 0$, requiring that the MutS-MutL has not arrived at the hemimethylated site. Using the method of images, this solution is given by

\begin{multline}
    p(x,t|x_0,x_{\textrm{meth}}) = \frac{1}{\sqrt{4 \pi D_{\textrm{SL}} t}} \left[ \exp \left( - \frac{\left( x - x_0 \right)^2}{4 D_{\textrm{SL}} t} \right)\right.  - \\ \left.\exp \left( - \frac{\left( x -\left(2 x_{\textrm{meth}} -  x_0 \right) \right)^2}{4 D_{\textrm{SL}} t} \right)  \right],
\label{eq:xmethboundpdf}
\end{multline} which represents the spatial probability density of the search at some time $t$ under the assumption that the search has not yet reached $x_{\textrm{meth}}$. 

\begin{eqnarray}
P(x < x_{\textrm{meth}},t)&=&\int_{- \infty}^{x_{\textrm{meth}}}  p(x,t|x_0,x_{\textrm{meth}}) \,\mathrm{d}x  \nonumber\\
&=& \erf { \left( \frac{x_{\textrm{meth}} - x_0}{\sqrt{4 D_{\textrm{SL}} t}} \right) }
\label{eq:firstPassageProb}
\end{eqnarray} thus gives the probability at time $t$ that a clamp has only searched positions $x_S$ such that $x_S < x_{\textrm{meth}}$. The probability that $x_{\textrm{meth}}$
 \textit{has} been searched, therefore, is 
 
 \begin{equation} 
 P^{(0)}(t) = 1 - P(x < x_{\textrm{meth}},t) = 1\!-\!\erf { \left( \frac{x_{\textrm{meth}} - x_0}{\sqrt{4 D_{\textrm{SL}} t}} \right) },
 \label{eq:foundprob_deriv}
 \end{equation} where the superscript $(0)$ indicates that the probability is for a non-dissociative search. Successful probability for a  dissociative search will be indicated by superscript $(*)$. 

\subsection{Association and Dissociation Event Stepping Simulation}\label{sec:adess}

In order to use the model described in Sec.~\ref{sec:model} to calculate successful search probabilities, we develop a Monte Carlo approach that samples from analytic one-dimensional diffusion probability distributions. This calculation breaks the problem of determining the overall successful search probability into the cumulation of the probabilities that each individual microscopic association identifies the hemimethylated site. Each individual probability can be determined analytically from the association lifetime and associated diffusion distributions if the distance between the position at which the clamps associate and the hemimethylated site is known.  For a given initial distance to the hemimethylated site, the subsequent hemimethylated site distances are determined by both the diffusion of the associated clamps and the diffusion of the dissociated clamps. In principle, this conceptual framework produces an analytic expression for the successful search probability involving iterative convolution integrals. In practice, however, this expression is too complex to be used to compute values directly. In particular, we found that the most straightforward way to calculate the many integrals over diffusion position and association lifetime probability distributions was to randomly sample from these distributions many times. Each set of random samples produces a probability of either 1 or 0 that the hemimethylated site was successfully reached, and the average of many of these sets gives the overall successful search probability. 

Another way to think of this iterative random sampling is to imagine that each set of random samples represents a path that the protein clamps can take along the DNA strand which results in either a successful or unsuccessful search. Each path occurs with a frequency proportional to its probability, and therefore setting the successful searches to 1 and the unsuccessful searches to 0 and taking the average of many such searches produces the successful search probability. 

The following algorithm is used to carry out this experiment, and will be called the association and dissociation event stepping simulation (ADESS):
\begin{enumerate}
	\item  The clamps start immediately adjacent to each other. We set the starting position of the clamps to $x_0=0$, the step counting index to $i=0$, and the elapsed time to $t_e=0$. Input a position to search for on a 1-D axis (designated the ``hemimethylated site" or simply ``$x_{\mathrm{meth}}$") representing its distance from the initial MutS-MutL association site on the dsDNA. Also choose a cutoff time, representing dissociation of the MutS clamp from the dsDNA. 
	\item Decide whether the adjacent clamps associate by sampling randomly from a uniform distribution between 0 and 1 and comparing the result to the input probability that adjacent clamps will associate, denoted $p_A$.
	\item If the clamps do not associate, go to step 7. If the cutoff time has been reached ($t_e\ge t_s$) mark the search as unsuccessful and go to to step 9.
	
	\item Randomly select, using the method of inverse transform, an association lifetime from the probability distribution given by
			 \begin{equation}
 		p_{\mathrm{assoc}}(t) = \tau_{\mathrm{A},\mu}^{-1} \exp(-t/\tau_{\mathrm{A},\mu}),
		 \label{eq:assocTime}
		 \end{equation}
		 where $\tau_{\mathrm{A},\mu}$ is the average microscopic association lifetime of the clamps. This represents the time for which the clamps are diffusing together during this association. Denote this time as $t_{\mathrm{assoc}}$ and increase the total elapsed time $t_e$ by $t_{\mathrm{assoc}}$.
		 
	\item Decide whether the hemimethylated site $x_{\mathrm{meth}}$ has been reached given the previous association position and lifetime by sampling randomly from a uniform distribution between 0 and 1 and comparing the result to the probability that the site has been reached, given by
	  \begin{equation}
 	P_{\mathrm{find}}(t_{\mathrm{assoc}}) = 1 - \text{erf} \left( \frac{x_{\mathrm{meth}} - x_i}{\sqrt{4 D_{\mathrm{SL},\mu} t_{\mathrm{assoc}}}} \right),
 	\label{eq:foundProb}
 	\end{equation} where $x_i$ is the previous association position and $D_{\mathrm{SL},\mu}$ is the diffusion rate of the associated clamps. Note that this is simply Eq.~(\ref{eq:foundprob_deriv}) evaluated at $t = t_{\textrm{assoc}}$. If the association time from the previous step brings the total time $t_e$ past the cutoff time, $t_{\textrm{assoc}}$ is taken to be the difference between the cutoff time and the time at which the current association began. This ensures that the final association finds the hemimethylated site with the proper probability.  
	
	\item 
	\begin{enumerate}
	\item If $x_{\mathrm{meth}}$ has been reached, the search is successful, so we proceed to step 9. 
	\item If $x_{\mathrm{meth}}$ has not been reached, use the previous association position $x_i$ and lifetime to randomly select the next dissociation position $x_{i+1}$ from the probability density function of Eq.~(\ref{eq:xmethboundpdf}) at $t = t_{\textrm{assoc}}$, with an additional normalization factor $C$ that ensures that the probability that $x_{\mathrm{meth}}$ has not been reached is 1 at time $t = t_{\textrm{assoc}}$. This factor is necessary because we have already determined in the previous step that the hemimethylated site has not been reached:
	\begin{eqnarray}
    \lefteqn{p(x_{i+1},t_{\textrm{assoc}}|x_i,x_{\textrm{meth}}) =}\hspace*{4mm}\nonumber\\
    &=&\frac{C}{\sqrt{4 \pi D_{\textrm{SL}} t_{\textrm{assoc}}}} \left[ \exp \left( - \frac{\left( x_{i+1} - x_i \right)^2}{4 D_{\textrm{SL}} t_{\textrm{assoc}}} \right)  -\right.\\
    &&\qquad\qquad\left.\exp \left( - \frac{\left( x_{i+1} -\left(2 x_{\textrm{meth}} -  x_i \right) \right)^2}{4 D_{\textrm{SL}} t_{\textrm{assoc}}} \right)  \right]\nonumber,
\label{eq:xmethboundpdf_tassoc}
\end{eqnarray} where
\begin{equation}
C = \frac{1}{\text{erf} \left[(x_{\mathrm{meth}} - x_i) / \sqrt{4 D_{\mathrm{SL},\mu} t_{\textrm{assoc}}} \right] }.
 \end{equation} 
    We increase $i$ by one to indicate that the $x_{i+1}$ determined here is the new position of the two newly dissociated clamps.
	\end{enumerate}
	\item Use the dissociation lifetime distribution
	  \begin{equation}
 	p_{\mathrm{dissoc}}(t) = \frac{x_d}{\sqrt{4 \pi D_{\mathrm{rel}} t^3}} \text{ exp} \left[- \frac{x_d^2}{4 D_{\mathrm{rel}} t} \right]
 	\label{eq:dissocTime}
 	\end{equation}
 	to determine how long the clamps remain dissociated (see Appendix~\ref{app:distributions}). Here, $x_d$ is as before the initial distance of the clamps following dissociation, and $D_{\mathrm{rel}}$ is the diffusion constant associated with the fluctuation of the distance between the clamps. Since each clamp is diffusing independently, the distance between them is also diffusing without bias in a particular direction. Denote this chosen lifetime $t_{\mathrm{dissoc}}$ and increment the total elapsed time $t_e$ by $t_{\mathrm{dissoc}}$.
	\item Using the lifetime chosen in the previous step $t_{\mathrm{dissoc}}$, select the next possible association position $x_{i+1}$ from the distribution of positions at which the relative position of the clamps returns to 0. This distribution is given by the solution to the unbounded diffusion equation with constant $D_ {\mathrm{CM}}$ associated with the diffusion of the ``center of mass'' of the dissociated clamps (see Appendix~\ref{app:distributions}). In particular,
	\begin{eqnarray}
 	\lefteqn{p_{\mathrm{return}}(x_{i+1} | x_{i}, t_{\mathrm{dissoc}})=}\nonumber\\ &=&  \frac{1}{\sqrt{4 \pi D_ {\mathrm{CM}}  t_{\mathrm{dissoc}}}} \text{ exp} \left[ -\frac{(x_{i+1} - x_{i})^2}{4 D_ {\mathrm{CM}}  t_{\mathrm{dissoc} }}\right]
  	\label{eq:reassocPosition}
 	\end{eqnarray}
 	Increase $i$ by one and return to step 2.
	\item Perform many such searches and assign a value of 1 to all those that are successful and 0 to those in which the cutoff time is reached without success. Take the average value of all of these searches to determine the successful search probability.  Divide the trials into 10 independent blocks of equal number of trials and calculate the search probability for each block to determine standard error. 
\end{enumerate}

\subsection{Determination of model parameters}
The model described above is written in terms of several microscopic parameters. In this section we will determine the values of these parameters. Some of these parameters can be calculated directly from experimentally measured values and are summarized in Tab.~\ref{table:calcParams}. For the remainder, we need to make reasonable assumptions about their values, summarized in Tab.~\ref{table:approxParams}. 

\begin{table}[ht]
\centering
\begin{tabular}{ |>{\raggedright}p{.25\columnwidth}|l|p{.23\columnwidth}|>{\arraybackslash}p{.23\columnwidth}| }
 \hline
 Parameter & Symbol & SL Value & SLH Value  \\
\hline
 Dissociated clamps relative position diffusion constant  & $D_{\mathrm{rel}}$ & $(1.5 \pm 0.6) \times 10^7$ bp$^2$/s & $(1.3 \pm 0.5) \times 10^6$ bp$^2$/s \\ 
 \hline	
  Dissociated clamps ``center of mass'' diffusion constant  & $D_{\mathrm{CM}}$ & $(7 \pm 3) \times 10^5$ bp$^2$/s & $(3.2 \pm 2.8) \times 10^5$ bp$^2$/s \\ 
 \hline
 MutS-MutL diffusion constant & $D_{\mathrm{SL},\mu} $, $D_{\mathrm{SL},\mu}$ & $(6 \pm 3) \times 10^4$  bp$^2$/s & $(8 \pm 5)\times 10^4$  bp$^2$/s \\
 \hline
 MutS-MutL association lifetime & $\tau_{\mathrm{A},\mu}$ & $0.03 \textrm{ s} \leq \tau_{\mathrm{A},\mu} < 30 \textrm{ s} $ & $0.03 \textrm{ s} \leq \tau_{\mathrm{A},\mu} < 30 \textrm{ s} $ \\
 \hline
 Distance from hemimethylated site & $x_{\mathrm{meth}}$ & $500$-$3000$ bp&$500$-$3000$ bp\\
 \hline
 Total search time & $t_s$ & $185\pm35$ s& $185\pm35$ s\\
 \hline
\end{tabular} \\
\caption{Model parameter values calculated from experimental observables.  The column labelled ``SL Value'' gives each value in the absence of MutH, whereas the column labelled ``SLH Value'' gives each value in the presence of MutH. }
\label{table:calcParams}
\end{table}
\begin{table}[ht]
\centering
\begin{tabular}{ |>{\raggedright}p{.41\columnwidth}|p{.21\columnwidth}|>{\arraybackslash}p{.31\columnwidth}| }
 \hline
Parameter & Symbol & Value  \\
\hline
Adjacent MutS-MutL association probability & $p_A$ & \rule{0pt}{\baselineskip}$10^{-4} \le p_A \leq 1$  \\
 \hline
MutS-MutL microscopic dissociation distance & $x_d$ & $1$ bp\\
 \hline
 MutS-MutL macroscopic dissociation distance & $x_M$ & $1000$ bp \\
 \hline
\end{tabular} \\
\caption{Estimated model parameter values.}
\label{table:approxParams}
\end{table}

The reason that the values of these parameters must be calculated or estimated rather than be measured directly is that the spatial resolution of the experiment is diffraction limited. Since the wavelength of visible light is on the order of hundreds of nm and the protein footprints are on the order of a few nm, the proteins interact on scales below the spatial sensitivity of the experiment.  Importantly, this implies that the clamps can {\em appear} to be associated with each other in the experiment, when they are closer than the spatial resolution of the experiment, even though they may or may not be in actual physical contact. In contrast, in our model we define the associated state  as the state in which the diffusion of the clamps is coupled, and the clamps have undergone some conformational change that allows them to interact more closely with the backbone and thus changes their diffusion rate. The dissociated state is the state in which the clamps are diffusing independently of each other. To avoid confusion, we will thus for the purposes of describing the calculation of model parameters from experimental observables denote the state in which the clamps are physically associated as ``microscopically associated,'' the state in which the clamps are physically dissociated but close enough that their positions are indistinguishable within the resolution of the experiment as ``proximate,'' and the state in which the clamps are physically dissociated and far enough away that their positions are distinguishable as ``macroscopically dissociated''.  In addition, we will use ``macroscopically associated'' to describe clamps that could be either ``microscopically associated'' or ``proximate'' and ``microscopically dissociated'' for clamps that could be either ``proximate'' or ''macrosopically dissociated''.

\subsubsection{Diffusion constants of individual clamps}

Since diffusion is scale invariant, there is no reason to believe that the
microscopic diffusion constants $D_S$ and $D_L$ of the individual clamps are different from their macroscopically measured values given in Tab.~\ref{table:expParams}. Rewriting
the diffusion of two clamps of different diffusion constants in terms of relative and ``center of mass'' coordinate yields $D_{\mathrm{rel}}= D_S + D_L$ for the diffusion of the relative coordinate and $D_ {\mathrm{CM}}=\frac{D_S D_L}{D_S + D_L}$ for the diffusion of the ``center of mass'' coordinate.

\subsubsection{Association lifetime and complex diffusion constant} 

The experiment measures the lifetime $\tau_{\mathrm{A,M}}$ and diffusion constant $D_{\mathrm{SL, M}}$ of macroscopically associated clamps (see Tab.~\ref{table:expParams}).  Since macroscopically associated clamps could be either microscopically associated or proximal, a macroscopic association event consists of a sequence of transitions between the microscopically associated state and the proximal state, where only after multiple excursions into the proximal state the clamps finally reach a distance that can be resolved in the experiment and thus reach the macroscopically dissociated state. Thus, the macroscopically measured lifetime $\tau_{\mathrm{A,M}}$ is an effective lifetime that integrates over many microscopic dissociation and re-association events, and the macroscopically measured diffusion constant $D_{\mathrm{SL, M}}$ is a temporal average of the diffusion constant of microscopically associated clamps $D_{\mathrm{SL, } \mu}$ and the diffusion constant of the center of mass of individual clamps $D_{\mathrm{CM}}$ during their excursions in the proximal state. 

In Appendix~\ref{app:lifetime}, we explicitly calculate how the macroscopically measured lifetime $\tau_{\mathrm{A,M}}$ that integrates over multiple microscopic dissociation and re-association events depends on the microscopic parameters of the model.  Solving this dependence for the microscopic association time yields
\begin{eqnarray}\label{assocTimeBody}
\tau_{\mathrm{A},\mu}&=&\frac{1}{\left[ \left(\langle N_A  \rangle - 1 \right) p_A+1 \right] } \left[\tau_{\mathrm{A,M}}  - \frac{x_{M} (x_{M} - x_d)}{D_{\mathrm{rel}}} \right] \nonumber\\ &\approx& \frac{\tau_{\mathrm{A,M}}}{\left[ \left(\langle N_A  \rangle - 1 \right) p_A+1 \right] },
\end{eqnarray}
where $p_A$ is the probability that adjacent MutS and MutL clamps will associate, and $\big \langle N_A \big \rangle = x_{M}/x_{d}$ is the number of times the clamps are in a microscopically adjacent state (making microscopic association possible) in a single macroscopic association. $x_d$ and $x_{M}$  are the microscopic and macroscopic association distances, respectively, so $x_d \ll x_M$.  The approximation in the second line of Eq.~(\ref{assocTimeBody}) holds for our specific values of the parameters as $\frac{x_{M} (x_{M} - x_d)}{D_{\mathrm{rel}}}\approx0.07\textrm{ s}$ and $\tau_{\mathrm{A,M}}\approx30 \textrm{ s}$. It implies that the time spent in the proximal state has a negligible contribution to the macroscopic association time due to the speed of the dissociated diffusion, even though the fact that a macroscopic association event consists of multiple microscopic association events {\em is} relevant as evidenced by the prefactor $[ \left(\langle N_A  \rangle - 1 \right) p_A+1]^{-1}$. Accordingly (see Appendix~\ref{app:diffConstant}), the excursions into the proximal state do not have a significant impact on the diffusion constant either due to their short durations. Thus,
\begin{equation}
D_{\mathrm{SL},\mu}  \approx D_{\mathrm{SL,M}}.
\end{equation}

\subsubsection{Distance from the nearest hemimethylated site}

\begin{figure}[ht]
    \centering
    \includegraphics[width=0.49\textwidth]{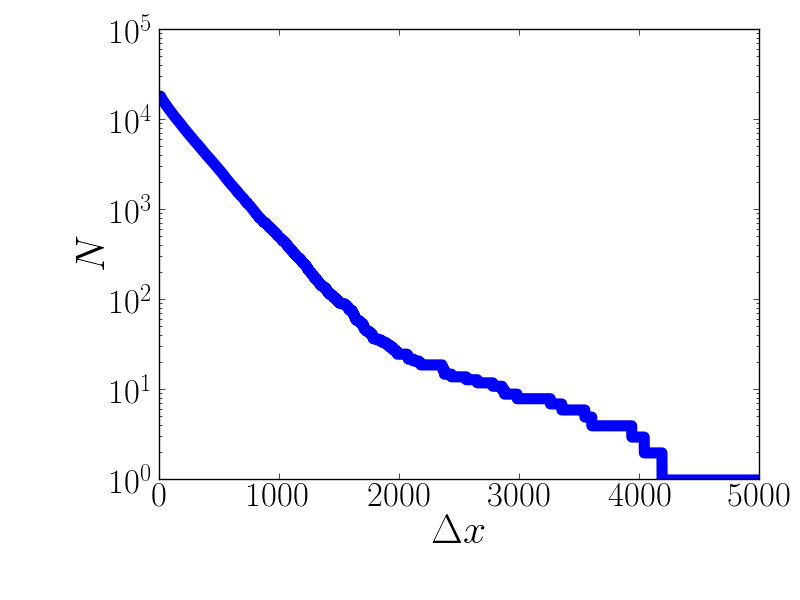}
    \caption{(color online) Distribution of hemimethylated site distances in the \textit{Escherichia coli} genome. For each separation on the horizontal axis, the vertical axis shows the number of adjacent GATC sites in the \textit{Escherichia coli} K-12 MG1655, NCBI RefSeq assembly GCF\_000005845.2 genome with at least that separation.}
    \label{fig:GATC_site_dist}
\end{figure}

In \textit{Escherichia coli}, hemimethylation occurs at GATC sites~\cite{lacks1977complementary, hattman1978sequence, geier1979recognition}.  Thus, the distance from a random location in the genome to the nearest hemimethylated site is governed by the distance distribution of adjacent GATC sites, shown in Fig.~\ref{fig:GATC_site_dist} for the genome of \textit{Escherichia coli} K-12 MG1655, NCBI RefSeq assembly: GCF\_000005845.2. While in 90\% of the cases, the distance between neighboring GATC sites is $500$ bp or less, the largest distances between adjacent GATC sites reach all the way to $5000$ bp. Since the ability to repair mismatches in the genome should depend on being able to identify the closest hemimethylated site even in the worst case scenario of being right in the middle of the two furthest separated GATC sites, we will report search probabilities over a range of $x_{\mathrm{meth}}=500-3000$ bp. 

\subsubsection{Total search time}

The search continues until either MutS or MutL dissociates from the the DNA. Since the experimentally determined MutS association lifetime $\tau_{\mathrm{S}} = 185 \pm 35$~s is much shorter than the experimentally determined MutL association lifetime $\tau_{\mathrm{L}} = 850 \pm 150$~s, the search time is limited by the MutS association lifetime and thus $t_s=\tau_{\mathrm{S}}$.

\subsubsection{Dissociation distances and association probability}

Unlike the microscopic association lifetime, microscopic diffusion constants, and the distance from hemimethylated sites, the dissociation distances $x_d$ and $x_M$ and the association probability $p_A$ are not determined by experimental observables, and thus cannot be calculated directly. Physical arguments, however, allow estimation of $x_d$ and $x_M$. In particular, the microscopic dissociation distance, i.e., the distance at which the clamps can be considered as independent, is on the order of $x_d \sim 1$ bp due to the base pair periodicity of the dsDNA free energy landscape. The macroscopic dissociation distance, i.e., the distance at which two clamps can be resolved in the experiment as being independent, is determined by the diffraction limit, and is expected to be about half the wavelength of the fluorescence. For red and green fluorescence, this distance is $x_M \sim 300 \textrm{ nm } \sim 1000$ bp.  

Similar physical arguments are unable to provide an estimate for the association probability $p_A$, but arguments can be made to set limits on this parameter.  As a probability, the upper limit on $p_A$ is evidently 1. Approximation of a lower limit is made possible by the assumption that $p_A \ge P_{\textrm{assoc, soln}}$, where $P_{\textrm{assoc, soln}}$ is the probability that a MutL in solution colliding with a DNA-bound MutS will associate. This assumption is plausible since there is only one dimension (namely rotation around the DNA) in which MutS and MutL clamps already associated with the DNA must align in order to associate with each other, rather than the three dimensions that must align when MutL is not already associated to the DNA. This assumption combined with published experimental results independent of the experiments in~\cite{Liu2016} suggests that the association probability must be greater than $10^{-4}$ (see Appendix~\ref{app:limitforpA}):
\begin{equation}\label{eq:pArange}
10^{-4}  \le p_A \leq 1
\end{equation}

\subsection{Validation of the ADESS approach}

\begin{figure}
\centering
	\includegraphics[width=0.49\textwidth]{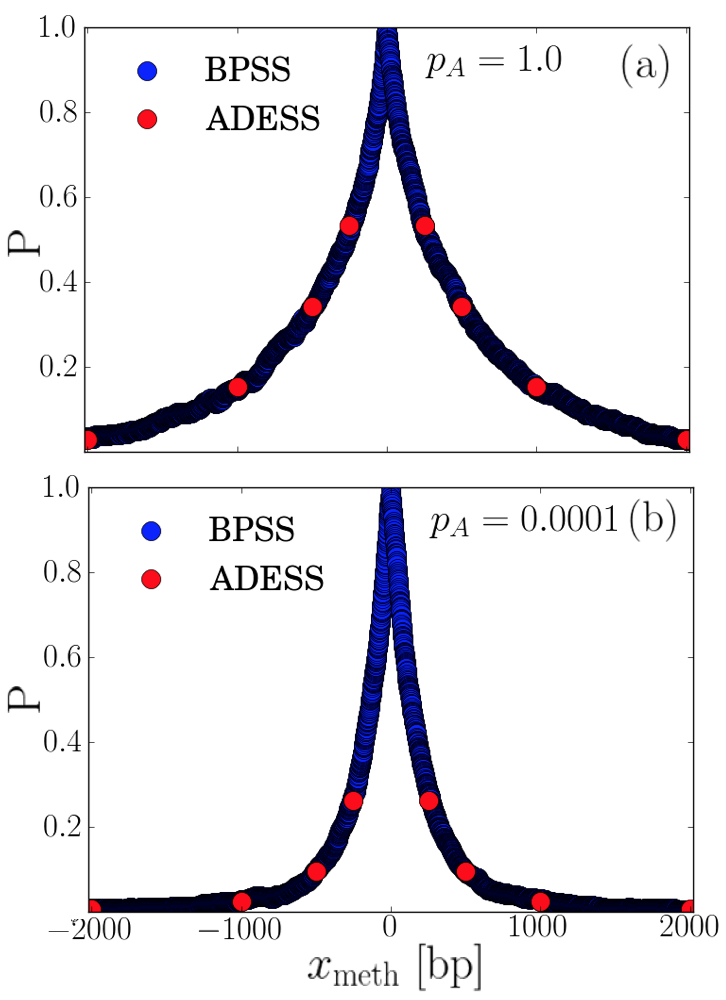}
\caption{(color online) Comparison between successful search probabilities calculated using BPSS and ADESS. (a) is calculated with $p_A = 1$ and (b) with $p_A = 10^{-4}$. The statistical uncertainty is smaller than the size of the symbols.}
\label{fig:dissCompProb}
\end{figure}
In order to validate the ADESS approach and the microscopic parameter calculation, we compare ADESS to a much more time consuming simulation that explicitly tracks the positions along the DNA and interactions of MutS, MutL, and MutS-MutL clamps. This simulation uses Gillespie's stochastic simulation algorithm~\cite{Gillespie1977} to choose a clamp and a direction in which to move it in every step (see Appendix~\ref{app:BPSS} for details). Each move has a step size of a single base pair; thus, we will denote this simulation approach as the base pair stepping simulation (BPSS). By counting those positions over which a MutS-MutL complex passed as having been searched, this simulation provides an alternative tool by which the successful search probability can be calculated. Since the BPSS approach follows every single diffusion step of the clamps, it becomes computationally unfeasible to obtain sufficient statistics for realistic values of the diffusion constants and we thus perform this validation for $D_S = 10^4$ bp$^2$/s, $D_{\textrm{SL}} = 10^3$ bp$^2$/s, and $D_{L} = 10^5$ bp$^2$/s, which are each about two orders of magnitude smaller than the actual experimentally determined diffusion constants.  Fig.~\ref{fig:dissCompProb} compares the search probability calculated using the BPSS approach and the search probability calculated using the ADESS approach and finds them to yield identical results within statistical error. 

\begin{figure}
\centering
	\includegraphics[width=\columnwidth]{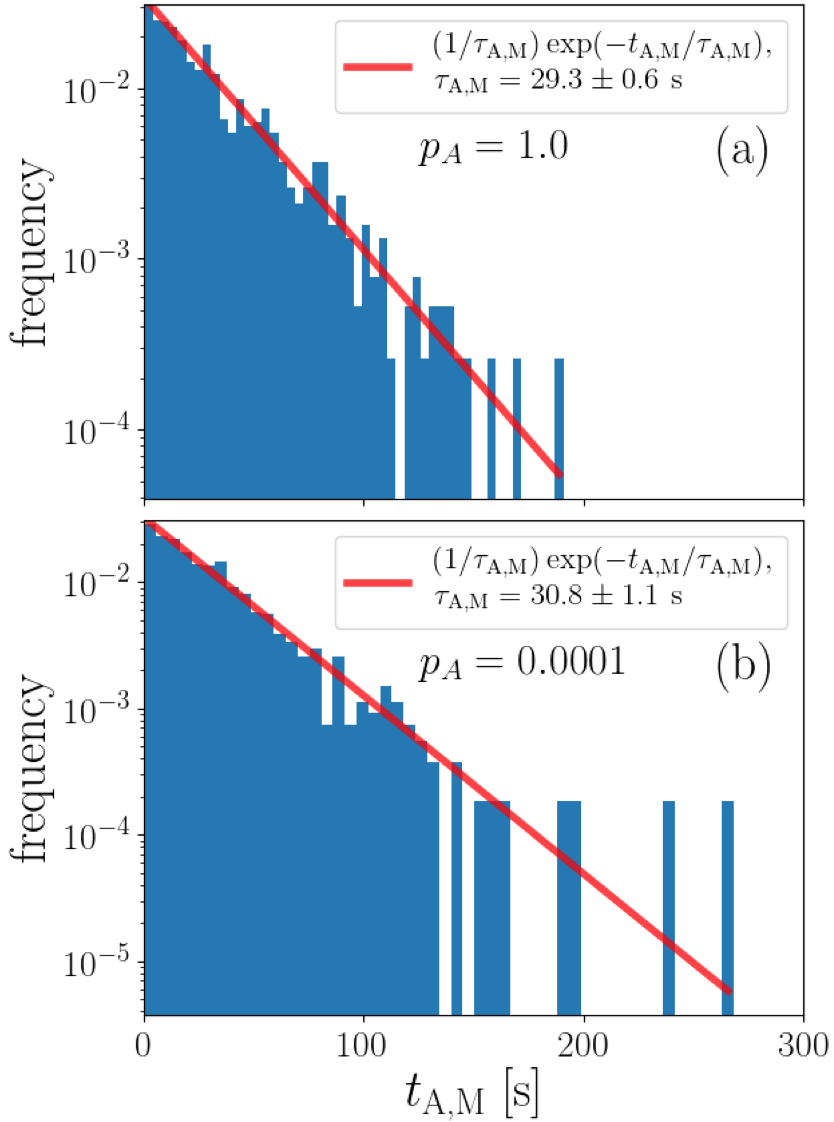}
\caption{(color online) Histogram of simulated macroscopic association times for (a)  $p_A = 1$ and (b) $p_A = 10^{-4}$. The line is given by $\tau_{\textrm{A,M}}^{-1} \exp{\big(-t_\mathrm{A,M}/\tau_{\textrm{A,M}}\big)}$, where $\tau_{\textrm{A,M}}$ is the average of the simulated macroscopic association times. This line therefore demonstrates that the association time probability decays exponentially with a decay constant consistent with the macroscopic association lifetime of \protect$\tau_{\mathrm{A,M}}\approx 30\pm3$ s measured experimentally in~\protect\cite{Liu2016}. The reported standard error of the decay constant is calculated by dividing the simulated data into 10 independent blocks and calculating the mean of each of them independently.}
\label{fig:GillespieMicroConfirmation}
\end{figure}

Additionally, the BPSS allows us to validate Eq.~(\ref{assocTimeBody}) for the microscopic association lifetime $\tau_{\mathrm{A},\mu}$ empirically. In particular, the BPSS approach lets us keep track of the distance between separate clamps and the times at which these distances occur. Using this feature, we calculate the time $t_{A,M}$ for which the clamps remain within the macroscopic association distance $x_M$ of each other, i.e., the time until they first reach the macroscopically dissociated state. Fig.~\ref{fig:GillespieMicroConfirmation} shows histograms of this time to reach the macroscopically dissociated state calculated from simulations that use the microscopic association lifetime calculated via Eq.~(\ref{assocTimeBody}).  We find that these simulated distributions accurately reproduce the experimentally measured macroscopic association lifetime $\tau_{\mathrm{A,M}}\approx 30\pm3$ s (see Tab.~\ref{table:expParams}), indicating that Eq.~(\ref{assocTimeBody}) correctly matches the microscopic association lifetime governing the multiple transitions between the microscopically associated and the proximal state to the macroscopic association lifetime observed in experiments. 

\subsection{Robustness of results to variation in estimated parameters}

Since several model parameters can only be estimated (see Tab.~\ref{table:approxParams}) we next determine how sensitive our model is to variations in these parameters.  The parameter with the largest uncertainty is the microscopic association probability $p_A$.  In order to gauge the sensitivity of the model to this parameter, we hold all other parameters constant at their values given in Tabs.~\ref{table:calcParams} and~\ref{table:approxParams} (both in the presence of, and the absence of, MutH) while varying the microscopic association probability over its entire potential range given in Eq.~(\ref{eq:pArange}).  Then, we numerically calculate the main observable of our model, namely the probability of a successful search, using the ADESS approach described in Sec.~\ref{sec:adess}.
\begin{figure}
\centering
\includegraphics[width=0.95\columnwidth]{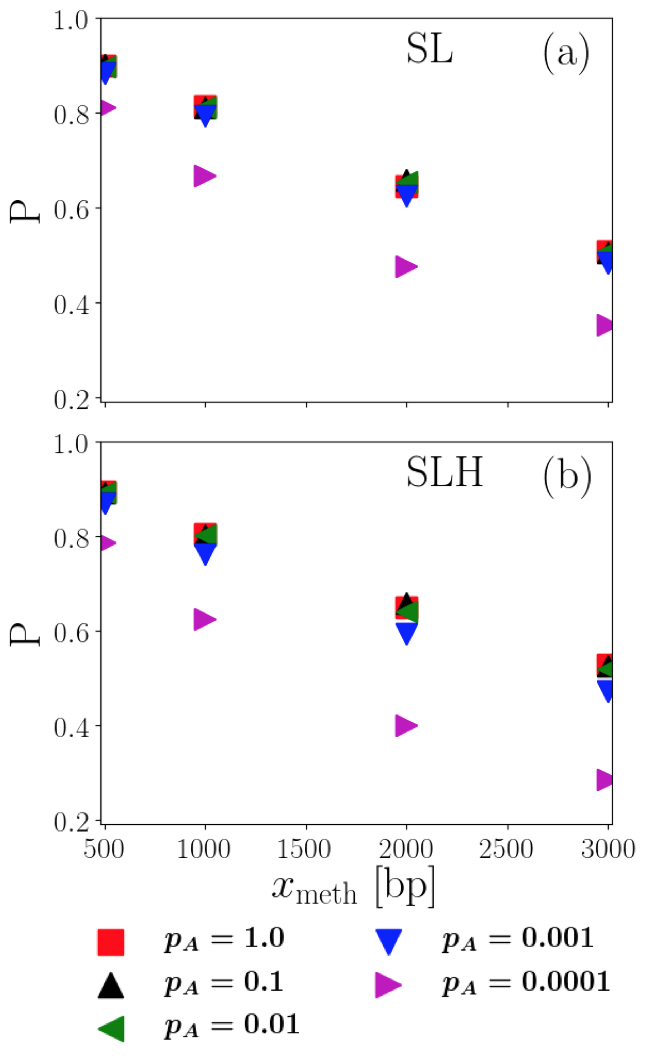}
\caption{(color online) Search probability as a function of search distance for different values of the association probability \protect$p_A$. (a) was calculated with non-MutH parameters, while (b) was calculated with MutH parameters. The statistical uncertainty is smaller than the size of the symbols.}
\label{fig:searchProbAssocProb}
\end{figure}
Fig. \ref{fig:searchProbAssocProb} shows the resulting search probabilities as a function of search distance $x_{\mathrm{meth}}$ for different values of the association probability $p_A$. We note that the successful search probability is largely independent of the microscopic association probability $p_A$ as long as $p_A\ge0.001$ and then drops significantly for $p_A=10^{-4}$. Since a significantly reduced search probability would be evolutionarily disadvantageous and our lower limit of $p_A\ge10^{-4}$ originated from a fairly generous ``worst case'' analysis (see Appendix~\ref{app:limitforpA}), we thus from here on focus on the range $0.001\le p_A\le 1$. In this range the search probability is largely insensitive to the value of $p_A$.

We note that naively it appears unintuitive for the overall search probability to be so insensitive to three orders of magnitude of variation in the probability that two adjacent clamps successfully form a complex. However, we would like to point out that the microscopic association probability $p_A$ appears in Eq.~(\ref{assocTimeBody}) for the microscopic association lifetime.  Thus, different values for the microscopic association probability $p_A$ yield different values for the microscopic association lifetime $\tau_{\mathrm{A},\mu}$ to keep the macroscopic association lifetime $\tau_{\mathrm{A,M}}$ consistent with its measured value.  The relative insensitivity of the search probability to the value of the microscopic association probability thus indicates that changes to the microscopic association lifetime compensate for the significant variation in microscopic association probabilities over three orders of magnitude.  This also explains the change in behavior at $p_A=0.001$.  Since the number of returns of the two clamps before final dissociation is $\langle N_A\rangle=1000$ for our parameters, the denominator $(\langle N_A\rangle-1)p_A+1$ in Eq.~(\ref{assocTimeBody}) is larger than one for $p_A\ge 0.001$ and asymptotes to one for $p_A<0.001$.  Thus, for $p_A\ge0.001$ the clamps go through multiple reassociation events before final dissociation, the lifetime of which compensates for the change in the microscopic association probability $p_A$. For $p_A<0.001$, the probability for even a single reassociation is becoming small and the microscopic association lifetime $\tau_{\mathrm{A},\mu}$ is locked to the macroscopic association lifetime $\tau_{\mathrm{A,M}}$, and is no longer able to compensate for changes in the association probability $p_A$.

\begin{figure}
\centering
	\includegraphics[width=\columnwidth]{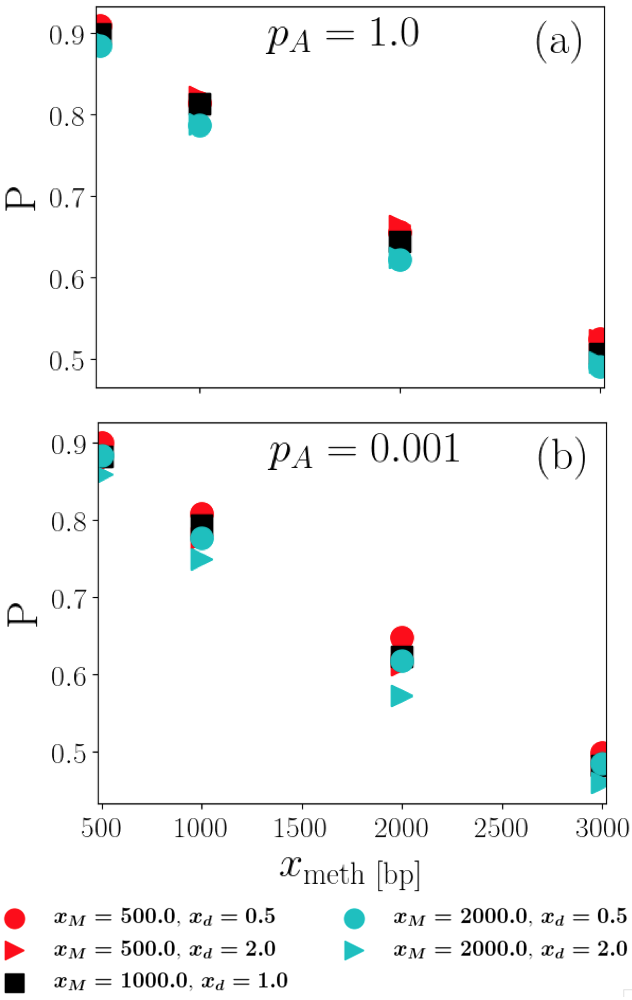}
\caption{(color online) Comparison of ADESS results for factor of two variations in the macroscopic and microscopic dissociation distances. (a) is calculated with $p_A = 1$ and (b) with $p_A = 0.001$. All results are shown for experimental parameters in the absence of MutH. The statistical uncertainty is smaller than the size of the symbols.}
\label{fig:varyAssumptions}
\end{figure}

Similar to our analysis of the sensitivity of the association probability $p_A$, we vary the values of the dissociation distances $x_d$ and $x_M$ by a factor of two in each direction to determine the sensitivity of the search probability to changes in these parameters at both limits of $p_A$. Fig.~\ref{fig:varyAssumptions} demonstrates that for $p_A=1$ and $p_A=10^{-3}$ variation of the dissociation distances $x_d$ and $x_M$ by a factor of two only introduces a relative difference of up to $13\%$. We thus conclude that the difference between the approximate and exact values of the dissociation distances $x_d$ and $x_M$ will not significantly affect our results.

\section{Dissociative search efficiency}\label{sec:results}

In this section we will systematically compare the efficiency of the dissociative search involving multiple dissociation-reassociation cycles of the two clamps with a non-dissociative search, in which the complex of the two clamps searches the DNA via simple diffusion.  The goal is to determine if the dissociative search observed in the experiments by Liu~\textit{et al.}~\cite{Liu2016} confers an evolutionary advantage of increased success probability over the simpler non-dissociative search. The successful search probability of the dissociative search is calculated numerically using the ADESS approach presented in Sec.~\ref{sec:adess}, while the successful search probability of the non-dissociative search is given analytically by Eq.~(\ref{eq:foundprob_deriv}). 

\subsection{Dissociative and non-dissociative searches result in similar single search efficiency for experimental diffusion constants}

\begin{figure}
\centering
	\includegraphics[width=\columnwidth]{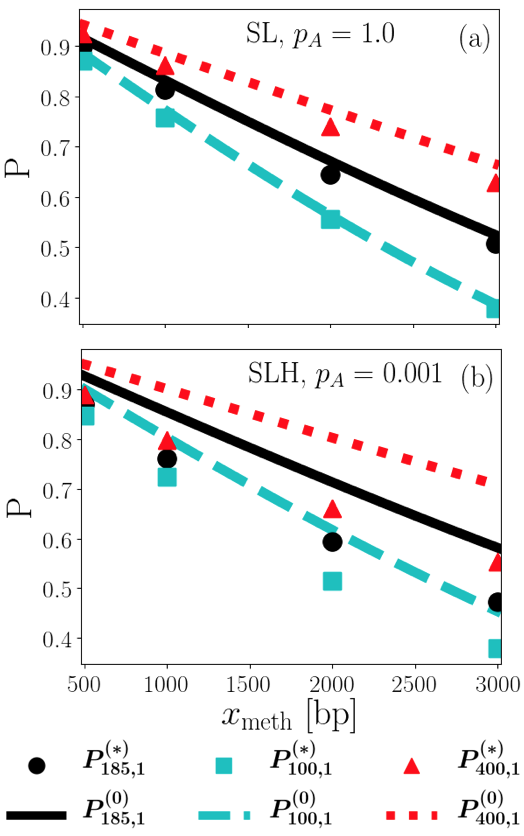}
\caption{(color online) Successful search probability of dissociative and non-dissociative searches as a function of distance $x_{\mathrm{meth}}$ from the hemimethylated site for different search times. Results in (a) use experimental parameters in the case where MutH is not associated with MutL and $p_A=1$ and (b) when MuH is associated with MutL and $p_A=0.001$. The statistical uncertainty is smaller than the size of the symbols. }
\label{fig:searchProb}
\end{figure}

Fig.~\ref{fig:searchProb} shows the successful search probability $P^{(*)}_{t_s, 1}$ of the dissociative search and $P^{(0)}_{t_s, 1}$ of the non-dissociative search for the experimentally determined values of the diffusion constants as a function of distance $x_{\mathrm{meth}}$ from the hemimethylated site. Here, the subscript $t_s$ indicates the search time in seconds, and the subscript $1$ indicates that the probability indicated is the success probability for only a single search. Probabilities are shown for various search times $t_s$ within roughly a factor of two from the experimental value of $185$ s in both directions.  The figure presents results for diffusion constants corresponding to the case where MutH is not associated with MutL and $p_A=1$ in (a) and for diffusion constants corresponding to the case where MutH is associated with MutL and $p_A=0.001$ in (b). These are chosen as the two extremes in terms of the differences between dissociative and non-dissociative searches, as the results for MutH parameters at $p_A=0.001$ and for non-MutH parameters at $p_A=1$ are in between the two cases shown.

Surprisingly, the non-dissociative search mechanism somewhat, but systematically, outperforms the dissociative mechanism for this choice of parameters, especially for the case of microscopic association probability $p_A=0.001$. In spite of these differences somwhat favoring the non-dissociative search mechanism, both search mechanisms result in sizeable successful search probabilities of at least $0.4$ for all parameter values explored here and thus both are likely to support successful DNA mismatch repair, in particular because it is likely that multiple searches occur during the mismatch repair process (see Sec.~\ref{section:multSearch}).

\subsection{Dissociative searches confer an advantage across a broad range of diffusion constants}

In the crowded \textit{in vivo} environment, diffusion is likely significantly slower (10-100 fold) than \textit{in vitro}~\cite{Konopka6115}. Additionally the diffusion constants, hemimethylated site distances, and association lifetimes of mismatch repair proteins may vary across organisms. In light of these observations, we next characterize the relative effect of the dissociative search mechanism across a wide range of possible diffusion rates. Although we only explicitly vary the diffusion rate, this can be seen as variation of the dimensionless combination $\sqrt{Dt}/x$ on which the probability depends (see Eq.~(\ref{eq:foundprob_deriv})). Thus, we effectively study variations in association time $t_s$ and hemimethylated site distance $x_{\mathrm{meth}}$ as well as diffusion rate.

\begin{figure}
\centering
	\includegraphics[width=\columnwidth]{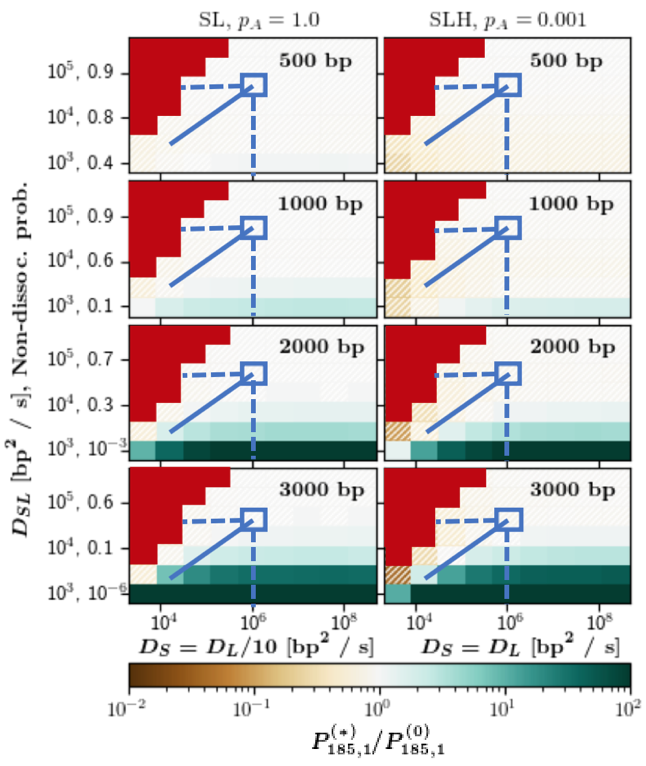}

\caption{(color online) Relative successful search probability as a function of diffusion constants for $D_S = D_L / 10\textrm{, } p_A = 1.0$ (left column) and $D_S = D_L\textrm{, } p_A = 0.001$ (right column). The former corresponds roughly to the case in which MutH is not present, and the latter corresponds roughly to the case in which MutH is present. The color scale indicates the ratio of ADESS dissociative and analytic non-dissociative probabilities, and is cut off at $10^2$ and $10^{-2}$ so that variations less than an order of magnitude are visible. Ratios greater than $10^{2}$ and less than $10^{-2}$ are set to $10^2$ and $10^{-2}$, respectively. The ratios less than one are hatched, while the ratios greater than one are solid. The square outlined indicates the order of magnitude of experimental diffusion constants, the possible \textit{in vivo} \textit{E. coli} diffusion constants are enclosed within the dashed lines, and the non-physical ($D_{\textrm{SL}} < D_S$) regions of the coefficient space are blocked out (in red). }
\label{fig:ratioScan}
\end{figure}

In order to characterize the effect of the dissociative search mechanism across many possible diffusion rates, times, and distances, we systematically vary diffusion rates and measure the relative advantage conferred by the dissociative mechanism. Fig.~\ref{fig:ratioScan} shows the relative probability $r$, defined as 
\begin{equation}
r \equiv P_{t_s,1}^{(*)} / P_{t_s,1}^{(0)}
 \end{equation}
 for $t_s=185$~s. The darkness of the color indicates the magnitude of the relative probability, and the squares that are brown and have hatching are those in which the dissociative mechanism lowers the successful search probability ($r< 1$), whereas the solid green squares indicate $r> 1$, i.e., areas of increased probability due to the dissociative search mechanism. To ensure that smaller differences are visible, relative differences $r> 100$ and $r< 1/100$ are set to $r=100$ and $r=1/100$, respectively. The slow, searching diffusion rate $D_{\mathrm{SL}}$ is varied along the vertical axis, while the fast diffusion rates $D_S$ and $D_L$ are varied along the horizontal axis. In order to restrict the plot to two dimensions, the ratio between the two fast rates is guided by experiment: either both rates are the same, or they differ by an order of magnitude, roughly corresponding to the situation in the presence and in the absence of MutH, respectively. The framed square corresponds to the \textit{in vitro} diffusion constants of \textit{E. coli} and the dashed lines enclose the range of diffusion constants that are smaller than the \textit{in vitro} diffusion constants, consistent with the \textit{in vivo} expectation~\cite{Konopka6115}.  The solid (blue) line indicates a reduction of the \textit{in vitro} diffusion constants by two orders of magnitude while maintaining the \textit{in vitro} ratio between $D_\textrm{SL}$ and $D_S$. It is important to note that although \textit{in vivo} diffusion rates for \textit{E. coli} are likely to fall within the region enclosed by the dashed lines, this may not necessarily be the case for other organisms. 

We find that differences between the search mechanisms are most significant for the largest distances from the hemimethylated site. Also, as expected, combinations of slow associated diffusion and fast dissociated diffusion are most favored by the dissociative mechanism (green/unhatched regions of the plot), whereas combinations of fast associated diffusion $D_{\mathrm{SL}}$ and slow dissociated diffusion $D_S$ and $D_L$ are least favored by the dissociative mechanism (brown/hatched regions of the plot). The latter case is often physically unrealistic since the associated clamps must diffuse more slowly than the individual clamps in order to interact with the DNA backbone and recognize the hemimethylated site. Accordingly the regions of the plot in which $D_{\mathrm{SL}} < D_S$ are blocked out in red, eliminating much of the space that would be disfavored by the dissociative mechanism. This renders the area favored by dissociation broad by comparison. 

The area which is most highly favored by the dissociative mechanism (the dark green region), however, occurs at very low single search success probabilities. While the relative probability increase is quite large ($\geq$ 100 fold increase in probability), it seems highly unlikely that a change from a non-dissociative success probability of, for instance, $10^{-6}$ to a dissociative success probability of $10^{-4}$ would cross some threshold probability below which failure of mismatch repair may negatively affect the organism. This point is emphasized by the inclusion of the absolute single non-dissociative search probability on the vertical axis (since this probability only depends on $D_{\textrm{SL}}$, it remains constant as one moves across the plot horizontally).  

\subsection{Multiple searches emphasize low probability single search differences}\label{section:multSearch}

\begin{figure}
\centering
	\includegraphics[width=\columnwidth]{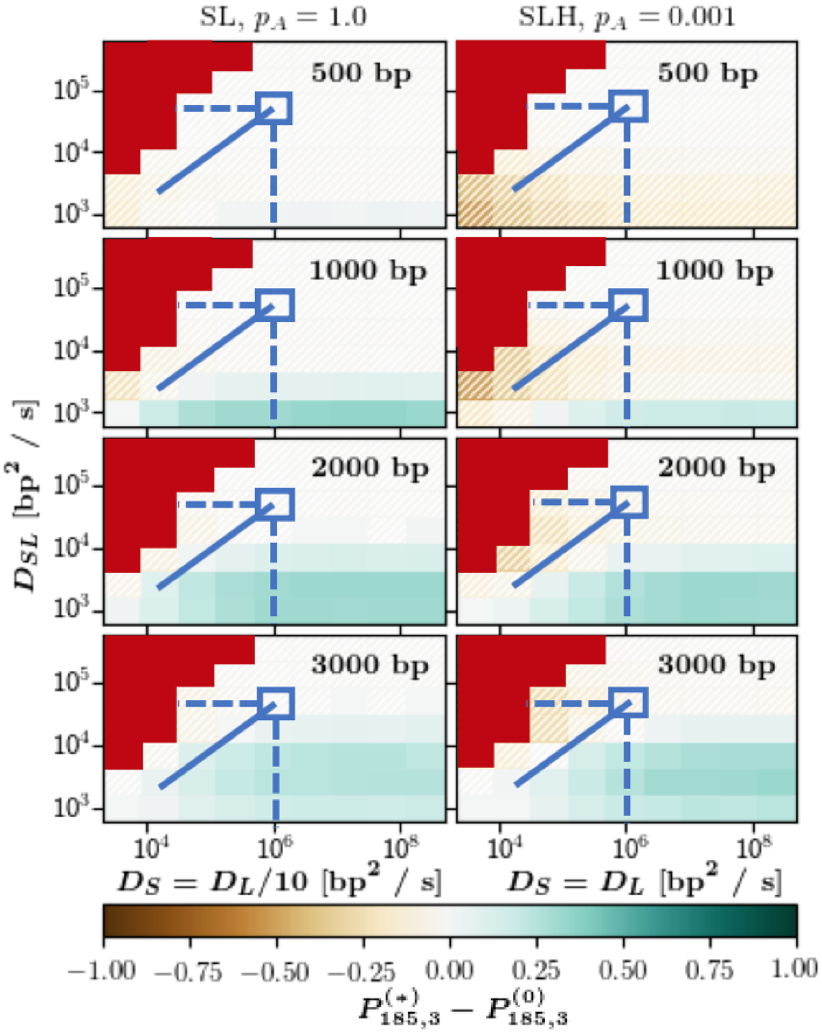}
\caption{(color online) diffusion constant space probability difference scan for searches by $n_s = 3$ protein complexes in the cases $D_S = D_L / 10\textrm{, } p_A = 1.0$ (left column) and $D_S = D_L\textrm{, } p_A = 0.001$ (right column). The former corresponds roughly to the case in which MutH is not present, and the latter corresponds roughly to the case in which MutH is present. The color scale indicates the absolute difference between the ADESS dissociative and analytic non-dissociative probabilities. Differences less than zero are hatched, while differences greater than zero are solid. The square outlined in blue indicates order of magnitude of experimental diffusion constants, the possible \textit{in vivo} \textit{E. coli} diffusion constants are enclosed within the dotted lines, and the non-physical ($D_{\textrm{SL}} < D_S$) regions of the coefficient space are blocked out (in red).}
\label{fig:diffScan_3}
\end{figure}

Data published by Acharya \textit{et al.}, Graham \textit{et al.}, and Hombauer \textit{et al.}~\cite{acharya2003,graham2018properties,hombauer2011visualization} suggest that the DNA mismatch repair process involves multiple MutS-MutL(-MutH) searches for the hemimethylated site. Thus, the cumulative probability for multiple low probability searches may result in a physiologically relevant success probability for the overall search process.  In order to approximate the effect of multiple searches, we need to calculate the probability that at least one search is successful. This quantity will be referred to as overall successful search probability. Although the proteins involved in separate searches are in principle able to interact with each other, accounting for these interactions is beyond the scope of this study. Instead, we hope to gain at least qualitative insight into the overall search probability under the assumption that the individual searches are independent. Under this assumption,
\begin{equation}
    P_{t_s,n_s} = 1 - (1 - P_{t_s,1})^{n_s}
\end{equation} where $P_{t_s,n_s}$ is the overall search probability, $P_{t_s,1}$ is the single search probability, and $n_s$ is the total number of searches. 

\begin{figure}
\centering
	\includegraphics[width=\columnwidth]{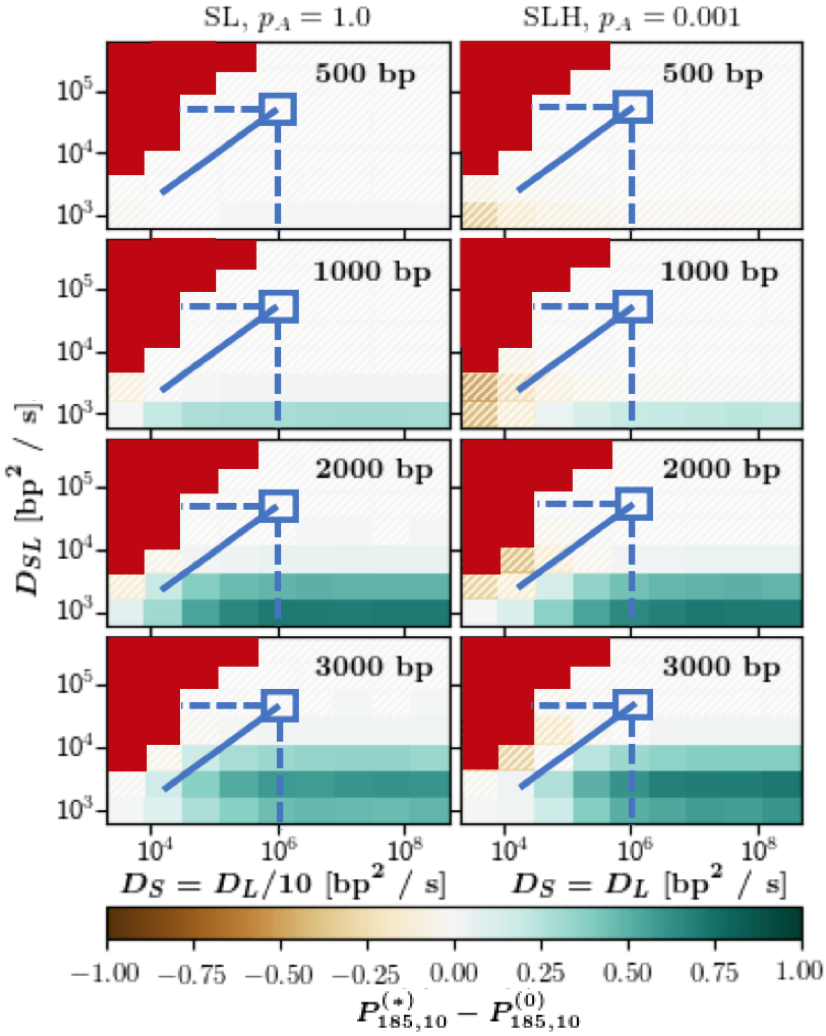}
\caption{(color online) diffusion constant space probability difference scan for searches by $n_s = 10$ protein complexes in the cases $D_S = D_L / 10\textrm{, } p_A = 1.0$ (left column) and $D_S = D_L\textrm{, } p_A = 0.001$ (right column). The former corresponds roughly to the case in which MutH is not present, and the latter corresponds roughly to the case in which MutH is present. The color scale indicates the absolute difference between the ADESS dissociative and analytic non-dissociative probabilities. Differences less than zero are hatched, while differences greater than zero are solid. The square outlined in blue indicates order of magnitude of experimental diffusion constants, the possible \textit{in vivo} \textit{E. coli} diffusion constants are enclosed within the dotted lines, and the non-physical ($D_{\textrm{SL}} < D_S$) regions of the coefficient space are blocked out (in red).}
\label{fig:diffScan_10}
\end{figure}

 Figs.~\ref{fig:diffScan_3} and~\ref{fig:diffScan_10} show diffusion space scans of
 \begin{equation}
     \delta P_{t_s,n_s} \equiv P_{t_s,n_s}^{(*)} - P_{t_s,n_s}^{(0)}
 \end{equation}
 indicated by the coloring/hatching for $n_S=3$ and $n_S=10$ searches, respectively. Note that in these figures difference between the two probabilities, rather than their ratio, is chosen to avoid overemphasizing large relative changes between two otherwise small probabilities. 
 
 As in Fig.~\ref{fig:ratioScan}, the physically unrealistic regions are blocked out, the probable region in which \textit{E. coli} diffusion constants reside are enclosed in the dotted lines, and the approximate \textit{in vitro} \textit{E. coli} diffusion constants are indicated by the blue square. Since probability differences are shown in the colormap, the probability of the non-dissociative search is omitted from the vertical axis. 
 
 Figs.~\ref{fig:diffScan_3} and~\ref{fig:diffScan_10} demonstrate that there is a much broader range of diffusion constants, and therefore hemimethylated site distances and association times, for which the dissociative search mechanism is beneficial for mismatch repair hemimethylated site searches as compared to pure diffusion. For $10$ searches, the absolute difference in probability approaches $\delta P_{185s,10} = 1$ for the cases in which dissociation is most favorable, whereas for $3$ searches the maximum difference in probability is more modest, with $\delta P_{185s,3} \approx 0.5$. The case with $3$ searches, however, exhibits a larger regime in which the dissociation mechanism is meaningfully beneficial.  
 
\section{Conclusions}\label{sec:conclusions}

Experiments by Liu~\textit{et al.}~\cite{Liu2016} observed repeated association and dissociation between MutS and MutL sliding clamps involved in identification of a hemimethylated site during DNA mismatch repair in \textit{E. coli}. This naturally raises the question if locally searching the DNA in the associated state and then quickly diffusing to a different location on the DNA when dissociated actually provides an advantage to the search process. Here, we model the dissociative search process, calculate the probability that searching DNA mismatch repair proteins successfully locate the hemimethylated site, and compare the success rate of this dissociative search to the success rate of a simple diffusive search. We find that both search mechanisms are highly efficient for the majority of observed hemimethylated site distances at measured \textit{in vitro} diffusion rates. Perhaps somewhat surprisingly, there is a slight disadvantage in terms of single search probability conferred by the dissociative search mechanism for searches at these \textit{in vitro} rates. We note, however, that there may be variation in diffusion rate, association lifetime, and hemimethylated site distance among different organisms and that it has been shown that \textit{in vivo} diffusion can be slower than \textit{in vitro} diffusion by one or two orders of magnitude~\cite{Konopka6115}. Accordingly, we studied the effect of the dissociative search mechanism across a large range of the parameter space of diffusion rates, association lifetimes, and hemimethylated site distances and found that the dissociative mechanism is either neutral or favorable in most cases.  We find the most significant advantages of the dissociative search in the parameter regime where the overall search probabilities (of both the dissociative and the non-dissociative searches) are very small. While successful search probabilities in the sub-percent range are probably not physiologically meaningful by themselves, we showed that they do become meaningful when taking into account that DNA mismatch repair includes multiple MutS initiated searches for the hemimethylated site, resulting in a physiologically relevant advantage of the dissociative search mechanism for large regions of the physically realistic parameter space.

It is important to emphasize that our treatments of multiple searches and \textit{in vivo} diffusion here are necessarily approximate. A more detailed treatment that accounts for the interactions between proteins that are initially involved in ``separate" searches may be a fruitful avenue for future research: in principle the base pair stepping simulation is capable of tracking more than two proteins, but the current computational cost is too high. Additionally, it is likely possible to expand the association and dissociation event stepping simulation to account for more than two proteins and the presence of other molecules on the DNA strand. In particular, the presence of other molecules on the DNA strand may provide a spatial constraint that prevents the occurrence of the of long-lived dissociation events that decrease the efficiency of the dissociative mechanism. Moreover, we have here assumed that the first encounter of a MutS-MutL complex with a hemimethylated site results in its recognition followed by an incision. If recognition of the hemimethylated site is stochastic itself, this will also reduce the overall search probability. Incorporating this effect into our approach and quantitating its consequences on the search probabilities of the dissociative and non-dissociative searches will be an interesting direction of future research.

Another potential avenue of study is the effect of a more physiological environment on the diffusion constants of the proteins. We note that the \textit{in vivo} diffusion constants are likely to be smaller than the measured \textit{in vitro} coefficients, but are not able to quantitatively predict the magnitude of this decrease. A study that determines the actual \textit{in vivo} diffusion constants of mismatch repair proteins could therefore be very useful. Similarly, determination of diffusion constants in systems other than \textit{E. coli} would be interesting. 

We note that in addition to its role in the search for a hemimethylated site, MutL acts as a processivity factor for the DNA helicase uvrD, resulting in the excision that is necessary for the progression MMR process~\cite{liu2019mutl}. It therefore could be the case that the observed dissociative mechanism is evolutionarily preferred  because the dissociation steps allow MutS to load multiple MutL proteins onto the strand, aiding in excision. This alternative hypothesis would be strengthened if further work determines that \textit{in vivo} search efficiency is not increased by the dissociative mechanism, although it is also possible that the dissociative mechanism serves a dual purpose: both increasing search efficiency and loading multiple MutL proteins onto the DNA strand.  

Beyond describing the specifics of the MutS-MutL search process, our approach in this paper is likely to be applicable to other diffusive processes along DNA in biology. For instance, Zessin \textit{et al.} observe a fast and slow diffusion rate of proliferating cell nuclear antigen (PCNA), which is a eukaryotic protein similar to a $\beta$ clamp that also forms a clamp structure during association with DNA~\cite{zessin2016}. Eukaryotes also exhibit three homologs to both MutS and MutL~\cite{Fishel2015}, combinations of which are likely to result in a variety of association/dissociation and diffusion parameters. In this case, the broad parameter space characterized by our analysis may provide insight into MMR in many organisms. 

Despite the work still necessary to fully understand the diffusive search process in DNA mismatch repair, we provide a broad characterization of the observed dissociative search mechanism along with a robust analytical and computational framework with which to study diffusion and interaction of protein clamps in DNA mismatch repair that can provide the basis for generalization to other sliding clamp systems in Biology.

\section{Acknowledgments}

This material is based upon work supported by the National Science Foundation under Grant No.~DMR-1719316 to RB and by the National Institutes of Health under Grant Nos.~GM129764 and~CA067007 to RF.


%\bibliography{clampsPaper}

\appendix
\section{Time and location of re-association}\label{app:distributions}

In this appendix we derive the probability densities for the time to reassociation and the reassociation location of two clamps once they have disassociated from each other.  These distributions are used in the ADESS approach to update the time and position after a microscopic excursion of the clamps.

\subsection{Independent diffusion of two sliding clamps}\label{indepDiff}

While the two clamps are diffusing independently, the state of the system is given by positions $x_S$ and $x_L$ of the MutS and the MutL clamp along the DNA, respectively.  The joint probability distribution for the two clamps follows the diffusion equation
   \begin{equation}
 \frac{ \partial p(x_{\mathrm{S}},\! x_{\mathrm{L}} | t)}{\partial t}\!=\!D_S \frac{ \partial^2 p(x_{\mathrm{S}},\! x_{\mathrm{L}} | t)}{\partial x_S^2}+\! D_L \frac{ \partial^2 p(x_{\mathrm{S}},\! x_{\mathrm{L}} | t)}{\partial x_L^2}.
 \end{equation}
By analogy to the Schr{\"o}dinger equation for a two-body quantum mechanical problem, this equation can be rewritten in terms of relative and ``center-of-mass'' coordinates. In particular, substituting
   \begin{eqnarray}
x_ {\mathrm{CM}}&\equiv&\frac{\frac{1}{D_S} x_S + \frac{1}{D_L} x_L}{\frac{1}{D_S} + \frac{1}{D_L}},\\
x_{\mathrm{rel}}&\equiv&x_S-x_L,\\
D_{\mathrm{CM}} &\equiv&\frac{D_S D_L}{D_S + D_L}\qquad\mbox{and}\\
 D_{\mathrm{rel}}&\equiv& D_S + D_L
 \end{eqnarray}
yields
\begin{eqnarray}
 \lefteqn{\frac{ \partial p(x_ {\mathrm{CM}}, x_{\mathrm{rel}} | t)}{\partial t} =}\quad\\
 &=&D_ {\mathrm{CM}} \frac{ \partial^2 p(x_ {\mathrm{CM}}, x_{\mathrm{rel}} | t)}{\partial x_ {\mathrm{CM}}^2} + D_{\mathrm{rel}} \frac{ \partial^2 p(x_ {\mathrm{CM}}, x_{\mathrm{rel}} | t)}{\partial x_{\mathrm{rel}}^2},\nonumber
\end{eqnarray}
which describes independent diffusion of the ``center of mass'' coordinate $x_{\mathrm{CM}}$ with diffusion constant $D_{\mathrm{CM}}$ and the relative coordinate $x_{\mathrm{rel}}$ with diffusion constant $D_{\mathrm{rel}}$.

\subsection{Time of reassociation}

In our model, the microscopic dissociation of the two clamps results in them being separated by the microscopic dissociation distance $x_d$. Since relative and center of mass position diffuse independently, the time to reassociation is the time the freely diffusing relative coordinate $x_{\mathrm{rel}}$ takes to reach $x_{\mathrm{rel}}=0$ when starting at $x_{\mathrm{rel}}=x_d$. This problem is mathematically equivalent to the problem of the associated clamps reaching the hemimethylated site $x_{\mathrm{meth}}$ after starting at some position $x_0$.  We can thus mirror image Eq.~(\ref{eq:firstPassageProb}) (since $x_{\mathrm{rel}}=0$ provides a left boundary for this problem while $x_{\mathrm{meth}}$ provided a right boundary in the context of Eq.~(\ref{eq:firstPassageProb})) and replace $x_0$ with $x_d$, $x_{\mathrm{meth}}$ with $0$, and $D_{\mathrm{SL}}$ with $D_{\mathrm{rel}}$ to obtain
  \begin{equation}
    P(t | x_{\mathrm{rel}} > 0) = \erf { \left( \frac{x_{d}}{\sqrt{4 D_{\mathrm{rel}} t}} \right) }
 \end{equation}
 for the probability that at time $t$ the two clamps starting at an initial distance of $x_d$ have not yet touched. The probability density associated with the return of the distance between the two clamps to $0$ from a distance of $x_d$ is therefore given by the negative derivative of this probability, i.e., 
\begin{eqnarray}
 p_{\mathrm{dissoc}}(t)&=&-\frac{\partial P(t | x_{\mathrm{rel}} < x_{\textrm{meth}})}{\partial t}\nonumber\\
 &=&\frac{x_d}{\sqrt{4 \pi D_{\mathrm{rel}} t^3}} \text{ exp} \left[- \frac{x_d^2}{4 D_{\mathrm{rel}} t} \right].
\end{eqnarray}

\subsection{Location of reassociation}
 
 Since at the time of reassociation the two clamps are at the same location, all we have to do to find the location of this event is to follow the motion of the center of mass coordinate $x_{\mathrm{CM}}$ during the excursion.  Since this is a free diffusion, the probability density for the location of the meeting point $x$ of the two clamps after a time $t$ given that they dissociated at some location $x_0$ is
\begin{equation}
 p_{\mathrm{return}}(x | x_0, t) = \\ \frac{1}{\sqrt{4 \pi D_ {\mathrm{CM}} t}} \text{ exp} \left[ -\frac{(x - x_0)^2}{4 D_ {\mathrm{CM}} t }\right].
 \end{equation}

\section{Microscopic Parameter Calculation}

The following are the full calculations used to determine the microscopic protein dynamics from experimental observables. In particular, we calculate the microscopic diffusion constant, $D_{\mathrm{SL}, \mu}$, and the microscopic association lifetime, $\tau_{\mathrm{A},\mu}$. The calculations of $P_{M}$ and $\tau (x)$ calculations closely follow~\cite{Bennaim2008}, a web published early draft of~\cite{Krapivsky2010}.

\subsection{MutS-MutL Association Lifetime}\label{app:lifetime}

First, we calculate the microscopic association lifetime.  Consider first the macroscopic association lifetime, which can be written as

\begin{equation}\label{avgTime}
\tau_{\mathrm{A,M}} = \tau_{\mathrm{A},\mu} \left[ \left( \langle N_A  \rangle - 1 \right) p_A+1 \right] + \tau_{R} (\big \langle N_A \big \rangle - 1) + \tau_{M}
\end{equation} where $N_A$ is the number of times the clamps are microscopically adjacent during a single macroscopic association, $p_A$ is the probability of microscopic association given that the clamps are adjacent, $\tau_{R}$ is the average time to return to the adjacent state, and $\tau_{M}$ is the average time to reach distance $x_{M}$ without returning to the adjacent state (i.e. the average time to macroscopic dissociation). Note that removing a single adjacent state from the factor multiplied by $p_A$ and multiplying it directly by $\tau_{\mathrm{A},\mu}$ ensures that there is at least one microscopic association in every macroscopic association. This must be true physically, since different diffusion rates are observed during macroscopic association.

Consider  $N_A$ for a complex starting in the aggregate state:
\begin{equation}
\begin{split}   
&P(N_A = 1) = P_{M} \\
&P(N_A = 2) = (1-P_M) P_{M} \\
&P(N_A = 3) = (1-P_M)^2 P_{M}  \\
&P(N_A) = (1-P_M)^{N_A - 1} P_{M}
\end{split}
\end{equation} where $P_{M}$ is the probability for a newly microscopically dissociated complex to go to $x_{M}$. Thus,
\begin{equation}
\big \langle N_A \big \rangle = P_M\sum\limits_{N_A = 1}^{\infty} N_A (1-P_M)^{N_A - 1}  = \frac{1}{P_{M}}.
\end{equation}

In order determine $P_M$ we first consider $P_M$ as a function of the distance between the clamps, which we will denote as $x$ for the remainder of this subsection to avoid the more cumbersome notation of $x_{\mathrm{rel}}$ used in the rest of the manuscript. Evaluation of this function at $x = x_d$ will give $P_M$. ($P_M(x)$ will refer to the probability to go to $x_M$ from some position $x$ without visiting $0$, while $P_M \equiv P_M(x_d)$ refers to the probability to go to $x_M$ from $x_d$.) Additionally, since the clamps diffuse with intermittent DNA contact, $P_M(x)$ will be calculated under the assumption that the distance between clamps diffuses continuously. This allows us to write
\begin{equation}\label{prob}
\begin{split}
&P_{M}(x) = \frac{1}{2} P_{M} (x + \delta x) + \frac{1}{2} P_{M} (x - \delta x) \\
&0 = \frac{P_{M} (x + \delta x) - 2 P_{M}(x) + P_{M} (x - \delta x)}{{\delta x}^2}
\end{split}
\end{equation}
and therefore 
\begin{equation}\label{probDiff}
\frac{{\partial}^2 P_{M}(x)}{\partial x^2} = 0
\end{equation}
with the boundary conditions
\begin{equation}
\begin{split}
&P_{M} (0) = 0 \\
&P_{M} (x_{M}) = 1.
\end{split}
\end{equation}
The unique solution of this differential equation is
\begin{equation}
P_{M} (x) = \frac{x}{x_{M}}
\end{equation}
and thus 
\begin{equation}\label{infProb}
P_{M} \equiv P_{M}(x_d) = \frac{x_d}{x_{M}}
\end{equation} where $x_d$ is the separation of the clamps immediately following dissociation. Therefore we conclude that 
\begin{equation}
\big \langle N_A \big \rangle = x_M / x_d.
\end{equation}

In order to compute the microscopic association lifetime $\tau_{\mathrm{A},\mu}$ from Eq.~(\ref{avgTime}), it is also necessary to compute the average return time $\tau_R$ and the average time $\tau_M$ to reach $x_{M}$. To this end, consider the average time $\tau(x)$ for the distance between the clamps to reach either $0$ or $x_{M}$ given that the starting distance is $x$:
\begin{equation}
\tau(x) = \sum\limits_{\textrm{paths}} t_p(x) P_p (x)
\end{equation} where $t_p(x)$ is the time for a path of length $x$ and $P_p(x)$ is the probability of such a path. Consideration of the effect of single infinitesimal time step $\delta t$ allows us to write
\begin{eqnarray}
\tau(x)&=&\sum\limits_{\textrm{paths}} t_p(x) P_p (x)\nonumber\\
&=&\sum\limits_{\textrm{paths}}\Big[ \frac{1}{2} t_p(x+\delta x) P_p(x+\delta x)+\\
&&\qquad\quad+\frac{1}{2} t_p (x-\delta x) P_p(x-\delta x) \Big] + \delta t\nonumber\\
&=&\frac{1}{2} \tau(x + \delta x) + \frac{1}{2} \tau (x - \delta x) + \delta t.\nonumber
\end{eqnarray}
Thus, division by the square of some small spatial step $\delta x^2$ yields
\begin{equation}
-\frac{2 \delta t}{{\delta x}^2} = \frac{\tau(x + \delta x) + \tau (x - \delta x) - 2 \tau (x)}{{\delta x}^2}.
\end{equation}
Therefore,
\begin{equation}
\frac{{\partial}^2 \tau(x)}{\partial x^2} = -\frac{2 \delta t}{{\delta x}^2} = - \frac{2}{D_{\mathrm{rel}}},
\end{equation}
where we write the right hand side in terms of the diffusion constant $D_{\mathrm{rel}} = D_S + D_L$. The boundary conditions 
\begin{equation}
\begin{split}
& \tau (0) = 0 \\
&\tau (x_{M}) = 0
\end{split}
\end{equation}
allow us to conclude
\begin{equation}
\tau (x) = \frac{x}{D_{\mathrm{rel}}} (x_{M} - x).
\end{equation}
We now write this quantity in terms of $\tau_{R}$ and $\tau_{M}$ as follows:
\begin{equation}\label{avgNonAssocTime}
\big \langle N_A \big \rangle \tau (x_d) =  \tau_{R} (\big \langle N_A \big \rangle - 1) + \tau_{M}.
\end{equation}
Thus, substitution into Eq.~(\ref{avgTime}) yields
\begin{equation}\label{avgTime2}
\tau_{\mathrm{A,M}} = \tau_{\mathrm{A},\mu} \left[ \left(\langle N_A  \rangle - 1 \right) p_A+1 \right]  + \big \langle N_A \big \rangle  \tau (x_d) 
\end{equation} Finally, we can conclude

\begin{equation}\label{assocTime}
\tau_{\mathrm{A},\mu} = \frac{\tau_{\mathrm{A,M}} - \big \langle N_A \big \rangle  \tau(x_d)}{\left[ \left(\langle N_A  \rangle - 1 \right) p_A+1 \right] } 
\end{equation} where $\big \langle N_A \big \rangle = x_M / x_d$.

\subsection{Microscopic Diffusion Constant}
\label{app:diffConstant}

Having computed the microscopic association lifetime, we turn our attention to the microscopic diffusion constant. During microscopic association, the observable quantity, that is, the  diffusion of the ``center of mass'' of the oscillating dissociative complex, is given by
\begin{equation}\label{DMSL}
D_{M, SL} = P_A D_{\mathrm{SL}, \mu} + P_D D_ {\mathrm{CM}},
\end{equation}
where $D_{\mathrm{SL},\mu}$ and $D_ {\mathrm{CM}}$ are the microscopically associated and dissociated complex diffusion rates, respectively, and $D_{M, SL}$ is the measured, macroscopic diffusion rate of the complex. $P_A$ and $P_D$ are the probabilities that the clamps are associated and dissociated, respectively. As argued in Sec.~\ref{indepDiff}, $D_ {\mathrm{CM}} = \frac{D_S D_L}{D_S + D_L}$. It follows that the quantity needed for the microscopic model, the microscopic diffusion constant, is given by
\begin{equation}\label{DA_D}
D_{\mathrm{SL, } \mu} = \frac{1}{P_A}(D_{M, SL} - P_{D} \frac{D_S D_L}{D_S + D_L})
\end{equation}

Since $D_{M,SL}$, $D_S$, and $D_L$ are measured experimentally, we only need to write $P_A$ and $P_D$ in terms of observable quantities to obtain a value for $D_{\mathrm{SL, } \mu}$. In order to do this, we observe that the probabilities that the proteins are microscopically associated and dissociated are given by the ratios of average time spent in an associated and dissociated state, respectively, divided by the sum of these times:
\begin{eqnarray}
P_A&=&\frac{p_A \tau_{\mathrm{A},\mu}}{p_A \tau_{\mathrm{A},\mu}+ \tau_{R}}\\
\label{PD}
P_D&=&\frac{\tau_{R}}{p_A \tau_{\mathrm{A},\mu} + \tau_{R}},
\end{eqnarray}
where $\tau_{\mathrm{A},\mu}$ is the microscopic association time, and $\tau_{R}$ is the average time to return to the adjacent state. $\tau_{\mathrm{A},\mu}$ is multiplied by the association probability, $p_A$, because there are $1/p_A$ returns with time $\tau_R$ for every microscopic association. Note that $\tau_{M}$ does not enter these equations. This is because the final walk from $x_\textrm{rel} = 0$ to $x_M$ has only a minor influence on the experimentally measured diffusion rate as $\tau_M$ represents only the last $\sim x_M^2 / D_\textrm{rel} \approx 0.1 \textrm{ } \mathrm{s}$ of the $\approx 30 \textrm{ } \mathrm{s}$ macroscopic association.  

Eq.~(\ref{avgNonAssocTime}) gives an expression for $\tau_{R}$ in terms of $\tau_{M}$, so in order to determine $\tau_{R}$ we must first compute $\tau_{M}$. Fortunately, we can calculate $\tau_{M}$ in a way that is analogous to the calculation of $\tau (x)$ in the previous section. Going back to a discrete picture, during a random walk that results in a separation distance $x = x_M$ before reaching $x = 0$, the first step after dissociation is from $x = x_d$ to $x = 2 x_d$. Thus,
\begin{equation}
\tau_{M} = \tau_{step} + \big \langle N_{x_d} \big \rangle \tau_{x_d, M} (2 x_d),
\end{equation}
where $\tau_{x_d, M}(x)$ is the average time for the distance between the clamps to reach either $x_d$ or $x_M$ and $N_{x_d}$ is the number of times the distance reaches $x_d$ before going to $x_{M}$. Modifying the calculation of $\tau (x)$ with the appropriate boundary conditions
\begin{equation}
\begin{split}
& \tau_{x_d, M} (x_d) = 0 \\
&\tau_{x_d, M} (x_{M}) = 0
\end{split}
\end{equation}
we find
\begin{equation}
\tau_{x_d, M} (x) = \frac{x-x_d}{D_{\mathrm{rel}}} (x_{M} - x)
\end{equation}
which yields
\begin{equation}
\tau_{M} = \frac{x_d^2}{ D_{\mathrm{rel}} }+  \frac{\big \langle N_{x_d} \big \rangle x_d }{D_{\mathrm{rel}}} (x_{M} - 2 x_d).
\end{equation}

Similarly, $\big \langle N_{x_d} \big \rangle$ can be computed in the same way that $\big \langle N_A \big \rangle$ was found earlier. In particular, 
\begin{equation}
\big \langle N_{x_d} \big \rangle = \frac{1}{P_{x_d, M}},
\end{equation}
where $P_{x_d, M}$ is the probability that the distance goes to $x_{M}$ before $x_d$ from distance $2 x_d$. 

Using Eqs.~(\ref{prob}) and~(\ref{probDiff}) with boundary conditions 
\begin{equation}
\begin{split}
&P_{M} (x_d) = 0 \\
&P_{M} (x_{M}) = 1
\end{split}
\end{equation}
we get
\begin{equation}
P_{x_d, M} = \frac{x - x_d}{x_{M} - x_d}.
\end{equation}
Finally, since we assume that the walk starts at $x = 2 x_d$,
\begin{equation}
\big \langle N_{x_d} \big \rangle = \frac{x_{M} - x_d}{x_d}.
\end{equation}
Appropriate substitutions and algebraic manipulations yield 
\begin{equation}
D_{\mathrm{SL, } \mu} = D_{\mathrm{SL,M}} - \delta \left(D_{\mathrm{CM}} - D_{\mathrm{SL,M}} \right)
\end{equation}
with 
\begin{eqnarray}
\delta&=&R_x R_\tau \frac{\left(2 - \frac{R_x}{1 - R_x}\right) \left(1 + \frac{R_x}{p_A (1 - R_x)}\right)}{\frac{1}{(1-R_x)}- R_\tau}\\
&\approx&2 R_x R_\tau \left(1+\frac{R_x}{p_A}\right)
\end{eqnarray}
where $R_x \equiv \frac{x_d}{x_M}  \sim 10^{-3}$, $R_\tau \equiv \frac{x_M^2}{\tau_{M,A} D_{\mathrm{rel}}} \sim 10^{-2}$ for the specific values of the parameters and the approximation in the second line holds since $R_x\ll1$ and $R_\tau\ll1$. In the following section we show that $10^{-4} \le p_A \leq 1$. For the experimental values of the parameters and $p_A=10^{-4}$ the correction $\delta(D_{\mathrm{CM}} - D_{\mathrm{SL,M}})$ is $\sim 50$ bp$^2/s$ $\sim 0.1 \% $ of $D_{\mathrm{SL,M}}$ and for $p_A = 1$, this correction is  $\sim 3$ bp$^2/s$ $\sim 0.01 \% $ of $D_{\mathrm{SL,M}}$. Thus,
\begin{equation}
    D_{\mathrm{SL, } \mu} \approx D_{\mathrm{SL,M}}.
\end{equation}

\subsection{Approximation of association probability lower limit}
\label{app:limitforpA}

The lower limit of the association probability can be calculated under the assumption that $p_A \ge P_{\textrm{assoc, soln}}$, where $P_{\textrm{assoc, soln}}$ is the probability that a MutL in solution colliding with a DNA-bound MutS will associate. As discussed in the main text of the paper, it should be easier for MutL and MutS to bind when they are both already somewhat aligned by their formation of clamp structures on the DNA.

The association probability $P_{\textrm{assoc, soln}}$ is given by the ratio 
\begin{equation}
    P_{\textrm{assoc, soln}} = k_{\textrm{on, exp}}/k_{\textrm{on, max}},
\end{equation}
where $k_{\textrm{on, exp}}$ is the experimental rate at which MutL associates with MutS on DNA from solution, and $k_{\textrm{on, max}}$ is the rate at which MutS and MutL collide (e.g. the diffusion limited rate). 

We first focus on the diffusion limited rate. The Smoluchowski equation yields an expression for the diffusion-limited rate constant for two uniform spheres~\cite{Smoluchowski}:
\begin{equation}
k_{\textrm{on, max}} = 4 \pi D R,
\end{equation}
where $D$ is the relative diffusion constant and $R$ is the reaction radius. 

Manelyte \textit{et al.} give the MutS Stokes radius as $R_{\textrm{S,S}} \sim 3 \text{ nm}$~\cite{10.1093/nar/gkl489}, and Grilley \textit{et al.} give the MutL Stoke radius as $R_{\textrm{S,L}} \sim 6 \text{ nm}$~\cite{grilley1989isolation}. Therefore $R \approx R_{\textrm{S,S}} +  R_{\textrm{S,L}}  \sim 10 \text{ nm}$. 

To determine the relative diffusion constant $D$, we use the measured MutS diffusion along the DNA strand, $D_S = 0.043 \pm 0.016 \text{ $\mu$m}^2 / \text{s}$, and the Stokes-Einstein diffusion of MutL in water at room temperature $D_{\mathrm{L},\text{ }\mathrm{soln}} = \frac{k_B T}{6 \pi \eta R_{\textrm{S,L}}} \approx 4 \times 10^{-11} m^2 / s \gg D_S$. Thus $D \sim 4 \times 10^{-11} m^2 / s$ and the diffusion limited on rate is
\begin{equation}
    k_{\textrm{on, max}} \sim 10^9 \textrm{ }\mathrm{M}^{-1} \mathrm{s}^{-1}.
\end{equation}

We can now turn to the experimental on rate. Liu \textit{et al.} do not measure this rate directly, but they do find the fraction $F_{\mathrm{SL}}$ of an ensemble of DNAs on which MutS-MutL complexes associate in equilibrium to be high enough to perform the experiment, i.e., a significant fraction of their constructs shows association of a MutL at their experimental concentration of MutL~\cite{Liu2016}. We thus choose $F_{\mathrm{SL}}=0.1$ as a conservative ``worst case'' estimate with $F_{\mathrm{SL}}\approx 1$ more likely. This, along with the known MutS dissociation constant with DNA, $K_{d,S} = 0.6\text{ } \mu M$~\cite{acharya2003coordinated} and the measured MutL off rate $k_{\textrm{off,L}} \sim 1 / \tau_{\mathrm{on,L}} \approx 1 / 850$ s can be used to estimate the desired on rate. The fraction of DNAs with MutS-MutL associated is given by 
\begin{equation}
F_{\mathrm{SL}} = \textrm{[SLDNA]} / \textrm{[DNA]} = \frac{k_{\textrm{on,L}} \textrm{[L]} \textrm{[SDNA]}}{k_{\textrm{off,L}}\textrm{[DNA]}}
\end{equation}
and thus
\begin{equation}
k_{on,L} = \frac{k_{\textrm{off,L}} F_{\mathrm{SL}} K_{d,S}}{[L][S]}.
\end{equation}
For the reported $[L] \approx 20$ nM and $[S] \approx 10$ nM
\begin{equation}
   k_{\textrm{on,L}} \sim 10^5\textrm{ } \mathrm{M}^{-1} \mathrm{s}^{-1}
\end{equation}
for the worst case estimate $F_{SL}=0.1$ and $k_{\textrm{on,L}} \sim 10^6 \text{ }\mathrm{ M}^{-1} \mathrm{s}^{-1}$ for $F_{SL}=1$. Thus we conclude that
\begin{equation}
 P_{assoc, soln} \sim 10^{-4}\textrm{ M$^{-1}$ s$^{-1}$}
\end{equation}
and therefore
\begin{equation}
10^{-4}  \le p_A \leq 1
\end{equation} 
which gets narrowed to $10^{-3}\leq p_A\leq1$ for $F_{SL}=1$.

\section{Base pair stepping simulation}\label{app:BPSS}

In this section, we discuss the base pair stepping simulation (BPSS), which is used to validate the ADESS in more detail. This simulation keeps track of the states of the system and uses Daniel Gillespie's ``stochastic simulation" algorithm to transition between states~\cite{Gillespie1977}. Briefly, each simulation state consists of an either dissociated or associated MutS and MutL, as well as their position(s) along a DNA strand. Transitions between states occur at rates determined by the microscopic parameters, which allow us to track the timing of each state relative to the beginning of the simulation.

The allowed transitions are as follows: 
	\begin{itemize}
	\item For the dissociated state
	 \begin{itemize}
	        \item with MutS and MutL adjacent
	       \begin{itemize}
	            \item MutS moves away from MutL with rate $k_S = D_S x_\textrm{step}^2$, where $x_\textrm{step} = 1$ bp is the simulation spatial step size
	           \item MutL moves away from MutS with rate $k_L = D_L x_\textrm{step}^2$
	           \item MutS and MutL form an associated complex with rate consistent with $p_A$, in particular $k_A = (k_S + k_L)\frac{p_A}{(1 - p_A)}$
	       \end{itemize} 
	       \item with MutS and MutL spatially separated
	            \begin{itemize}
	                \item MutS moves away from MutL with rate $k_S = D_S x_\textrm{step}^2$
	                \item MutL moves away from MutS with rate $k_L = D_L x_\textrm{step}^2$
                    \item MutS moves toward  MutL with rate $k_S = D_S x_\textrm{step}^2$
	                \item MutL moves toward MutS with rate $k_L = D_L x_\textrm{step}^2$
	           \end{itemize} 
	        \end{itemize}
	\item For the associated state
	 \begin{itemize}
	 \item[]
	           \begin{itemize}
	           \item Move left or right with rate $k_\mathrm{SL} = D_{\mathrm{SL}} x_\textrm{step}^2$ each
	           \item Dissociate with rate $k_D = 1/\tau_{\mathrm{A},\mu}$. After dissociation, the bases are placed $1$ bp apart. This is achieved by moving one protein by $1$ bp away from the last complex position and leaving the other protein at the last complex position. MutS is moved with probability $k_S / (k_S + k_L)$, and MutL is moved with probability $k_L / (k_S + k_L)$. 
	           \end{itemize}
	 \end{itemize}
	\end{itemize}

In order to calculate observables with this simulation, we start with the proteins in an associated state at position $0$ and track their positions along the strand as a function of time.  Assuming that the associated complex searches every position that it passes, the fraction of simulations in which the complex has passed a specific position in the given amount of time is the overall successful search probability at that position. Additionally, we can use the distance that separates dissociated MutS and MutL clamps at a given time to calculate the macroscopic association time. In particular, the time at which the distance between the clamps reaches $x_M$ is recorded for each simulation and then the average of these times is used to calculate a decay constant, as in Fig.~\ref{fig:GillespieMicroConfirmation}.

\end{document}